\documentclass[11pt,a4paper]{article}
\pdfoutput=1

\usepackage{graphicx}
\usepackage{jcappub}

\usepackage{amsmath,amssymb,amsfonts}
\usepackage{color}
\usepackage{hyperref}

\graphicspath{
  {./}
  {./figures/}
}

\title{Gravitational waves from non-Abelian gauge fields at a tachyonic transition}

\author[a]{Anders Tranberg,}
\author[b]{Sara T\"ahtinen,}
\author[b]{David J. Weir,}
\affiliation[a]{Faculty of Science and Technology,
  University of Stavanger,\\
  N-4036 Stavanger,
  Norway}
\affiliation[b]{Department of Physics and Helsinki Institute of Physics,\\
  P.O.~Box 64,
  FI-00014 University of Helsinki,
  Finland}
\emailAdd{anders.tranberg@uis.no}
\emailAdd{sara.tahtinen@helsinki.fi}
\emailAdd{david.weir@helsinki.fi}


\abstract{We compute the gravitational wave spectrum from a tachyonic
  preheating transition of a Standard Model-like SU(2)-Higgs system.
  Tachyonic preheating involves exponentially growing IR modes, at
  scales as large as the horizon. Such a transition at the
  electroweak scale could be detectable by LISA, if these
  non-perturbatively large modes translate into non-linear dynamics
  sourcing gravitational waves.  Through large-scale numerical
  simulations, we find that the spectrum of gravitational waves does
  not exhibit such IR features.  Instead, we find two peaks
  corresponding to the Higgs and gauge field mass, respectively.  We
  find that the gravitational wave production is reduced when adding
  non-Abelian gauge fields to a scalar-only theory, but increases when
  adding Abelian gauge fields. In particular, gauge fields suppress
  the gravitational wave spectrum in the IR.  A tachyonic transition
  in the early Universe will therefore not be detectable by LISA, even
  if it involves non-Abelian gauge fields.}

\begin{document}

\begin{flushright}
  HIP-2017-09/TH
\end{flushright}

\maketitle

\section{Introduction}
\label{sec:intro}

The ground-breaking direct detection of gravitational waves
\cite{Abbott:2016blz} gives promise that cosmological sources may also
be detectable in the foreseeable future. One mission with scope to
look for such sources is LISA, due for launch in
2034~\cite{Audley:2017drz}. The primary contenders for detection are
gravitational waves from inflation (see for instance
\cite{Bartolo:2016ami}), from cosmic defects (see for instance
\cite{Binetruy:2012ze}) and from bubble collisions at a first order
phase transition (see for instance \cite{Caprini:2015zlo}). The latter
can in turn be connected to the creation of the cosmological baryon
asymmetry if the phase transition in question is the electroweak one,
at a temperature of around 100 GeV~\cite{Morrissey:2012db}.

These processes are favoured observationally by LISA, because the
scale of the dynamics is not primarily set by the microscopic
properties of the system, such as particle masses. The long-wavelength
behaviour of a system can play a significant role as well. For a
first-order phase transition, this might correspond to the radius of
bubbles of the new phase, which may grow to near-horizon scales.
Similarly, cosmic strings potentially extend to the horizon and beyond
and could give observable signals. In contrast, frequencies
corresponding to electroweak mass-scales in the early Universe are
much too high to be detectable by LISA, even when redshifted to the
present epoch.

Another phenomenon with similar features arises when symmetry breaking
is triggered at low temperature. Rather than a thermal phase
transition, a spinodal (or tachyonic) decomposition occurs, whereby
all momentum modes of the field with $|k|$ smaller than some mass
scale $\mu$ grow exponentially in time. The UV effective cut-off $\mu$
is fixed by the microscopic physics of a given model, but the active
IR part of the spectrum stretches all the way to $|{\bf k}|=0$, or in
a expanding Universe, to the Hubble scale. There is therefore hope
that the large-amplitude momentum range in the associated
gravitational wave spectrum may overlap with the one probed by
LISA. We will investigate this here, for the case where the system
includes non-Abelian gauge fields.

For all these different phenomena mentioned, numerical simulations are
employed to compute the spectrum and strength of the gravitational
wave signal.  This is necessary, as the sources involve inhomogeneous,
non-perturbative field
dynamics~\cite{Dufaux:2007pt,GarciaBellido:2007af,GarciaBellido:2007dg,Dufaux:2010cf,Figueroa:2012kw,Hindmarsh:2013xza,Hindmarsh:2015qta,Figueroa:2016ojl,Hindmarsh:2017gnf}. Reheating
at the end of inflation is typically modelled by one or more
(self-)interacting scalar fields, which may or may not be coupled to
gauge fields. For baryogenesis at a first order thermal phase
transition, multiple fields are in play, but from the point of view of
gravitational wave creation, these are likely well modelled by an
ambient fluid, interacting with the Higgs field
wall~\cite{Hindmarsh:2013xza,Hindmarsh:2015qta,Hogan:1986qda}.

\subsection{Tachyonic transitions}

Spinodal decompositions are well-studied in condensed matter systems,
but in a cosmological context, they are traditionally associated with
hybrid inflation. This involves an inflation $\sigma$, coupled to a
second scalar $\phi$. As $\sigma$ slow-rolls below a certain critical
value, the effective mass parameter of the $\phi$ field becomes
negative, and the transition is triggered. As a result, the slow-roll
stage also ends, allowing for graceful exit from inflation.

But a tachyonic transition may arise in a wide variety of settings, as
long as the dynamics of one field triggers the symmetry breaking of
another, at low temperature. Examples of this include small- or
large-field inflationary models, where slow-roll inflation has ended
of its own accord, long before the symmetry breaking transition is
triggered \cite{german,bartjan,ross}; scenarios including spectator
fields playing the role of $\sigma$, rolling from some non-zero
initial condition set by the inflationary stage~\cite{enqvist}; or
indeed cases where the $\sigma$ field is itself undergoing a symmetry
breaking transition. There are also models where the scalar potential
of a single field $\phi$ is such that a first order ``tunnelling''
occurs followed by tachyonic roll-down into the zero-temperature
minimum \cite{konstandin,vonHarling:2017yew}.

In the present work, we wish to investigate the IR properties of the
gravitational wave (GW) spectrum from such a transition, and also to
stay agnostic about the specific triggering mechanism and embedding in
a UV theory. We will therefore model the quench in terms of a
time-dependent mass. Writing for future convenience in terms of a
complex scalar, we have for the second field $\phi$
\begin{eqnarray}
V(\phi)= V_0 +\mu^2_{\rm eff}(t)\phi^\dagger\phi + \lambda (\phi^\dagger\phi)^2,
\end{eqnarray}
where $\mu_{\rm eff}^2(t)$ is a model-dependent function of time
(given by the motion of the inflaton or spectator field, or even of temperature). It is assumed to
evolve from being positive to being negative, thereby triggering the
symmetry breaking transition. We can model it as\footnote{Our quench
  speed $u$ is equivalent to the quantity $V_c$ in
  \cite{Dufaux:2008dn}.}
\begin{eqnarray}
\label{eq:mu2}
\mu^2_{\rm eff}(t) = \mu^2\left(1-\frac{2t}{\tau_q}\right), \qquad u = -\left.\frac{1}{2\mu^3}\frac{d\mu_{\rm eff}^2(t)}{dt}\right|_{\mu_{\rm eff}=0}=\frac{1}{\mu\tau_q},
\end{eqnarray}
with the understanding that the time dependence applies to the
interval $0\leq t \leq \tau_q$, and for $t>\tau_q$, $\mu_{\rm
  eff}=-\mu^2$.  Matching to, for instance, a quartic ``portal''
coupling model $\xi^2\sigma^2\phi^\dagger\phi$, we could imagine
writing
\begin{eqnarray}
\label{eq:portal}
\mu^2_{\rm eff}(t) = (\xi^2 \sigma^2-\mu^2),\qquad u = -\left.\frac{1}{2\mu^3}\frac{d\mu_{\rm eff}^2(t)}{dt}\right|_{\mu_{\rm eff}=0}=-\frac{\xi \dot{\sigma}_c}{\mu^2},
\end{eqnarray}
so that $\tau_q\simeq -\mu/{\xi\dot{\sigma}_c}$, where the subscript
$c$ refers to the time of the quench $\sigma_c = \mu/\xi$.

Such a transition results in exponentially growing field modes with
$|{\bf k}|\leq \mu$ \cite{Felder:2000hj}. The subsequent
redistribution of the initial potential energy in $V_0$ is a highly
effective preheating mechanism. In a given model, the additional
kinetic energy of the $\sigma$ field, and possible resonances
(resonant preheating) must be considered (see for instance
\cite{Mou:2017xbo} in the context of baryogenesis).

The process of preheating through a spinodal transition is a violent
and inhomogeneous process, and produces gravitational waves
\cite{Dufaux:2008dn}. Even though the characteristic scales of the
transition (typically $\mu$ and $\tau_q$) are of the order of a GeV or
more, and hence way beyond the sensitivity range of detectors such as
LISA, the spectrum potentially extends in the IR to the scale of the
horizon.

Previous simulations of scalar fields
only~\cite{Dufaux:2007pt,GarciaBellido:2007dg} have shown that
gravitational waves are indeed produced in such a transition, but that
the spectrum tends to peak around the scale of the particle masses. To
the IR of this peak, there is first a $\propto k$ behaviour and then
$\propto k^3$. This in spite of there being high occupation numbers in
the field modes all the way to $k=0$. For scalar fields, this is
perhaps not unexpected given the form of the relevant source term (see
below), but still disappointing. However, since these models are not
directly connected to known physics (such as the Standard Model),
there is some freedom in choosing the couplings and energy scale,
including the quench speed $u$. In this way, one may construct models
whose signal approaches the LISA-detectable region.

Whereas the $\sigma$ field is often taken to be a gauge singlet, the
second field $\phi$ need not be. Guided by the situation in the
Standard Model, it is natural to expect both Abelian ($\mathrm{U}(1)$)
and non-Abelian ($\mathrm{SU}(N)$) gauge fields to couple to such a
``Higgs'' field, and participate in the preheating mechanism
\cite{Skullerud:2003ki,Lozanov:2016pac}. One may even entertain the
notion that the second field {\it is} the Standard Model Higgs
field. In this case, a fast spinodal transition from zero temperature,
may be responsible for the baryon asymmetry of the
Universe~\cite{GarciaBellido:1999sv,Krauss:1999ng,Copeland:2001qw,Tranberg:2003gi}.

Reported investigations of GWs from tachyonic transitions in
gauge-Higgs models consider the case where the gauge group is Abelian
\cite{Dufaux:2010cf}. In that case the peak of the spectrum still
stays in the UV, corresponding to the particle masses (scalar and
gauge). Again, some freedom in the choice of parameters means it is
possible to shift the peak amplitude and position towards the LISA
detection region.

A crucial difference between the U(1)-Higgs and SU(2)-Higgs
transitions is that in breaking the U(1) gauge symmetry, topological
defects are created in the form of Abelian Higgs strings. A
substantial literature exists on the late-time production of GWs from
such networks of cosmic strings, including large-scale numerical
simulations (see for instance the recent~\cite{Hindmarsh:2017qff}). But also
for short times, cosmic strings seem to give a contribution distinct
from that due to the tachyonic dynamics itself \cite{Dufaux:2010cf}.

Apart from it being a close analogue of the Standard Model, one upshot
of investigating the SU(2)-Higgs model is that such topological
defects are absent\footnote{See however \cite{sexty} for an analysis
  of transient local energy blobs related to ``textures'', ``half-knots''
  and oscillons in that context.}. This allows us to focus on the
GW-production from the spinodal growing IR modes themselves. Since
non-Abelian fields self-interact strongly, such a contribution could a
priori could be significant.

In the present work we will compute the
spectrum of gravitational waves, as a function of the mass scale $\mu$
and the quench time $\tau_q$. For most of our simulations, we will be
conservative and assume a Standard Model-like theory, by setting the
Higgs and gauge (self-)interaction couplings $\lambda$ and $g^2$ equal
to their Standard Model values. Hence, when the scale $\mu$ is around
100 GeV, our results apply directly to a Standard Model
transition. When $\mu\gg 100 $ GeV, the theory is a specific
realisation of a generic SU(2)-Higgs system.

When appropriate, we will compare to a scalar-only theory as well as
simulations of an Abelian U(1)-Higgs model. We stress again that such
comparisons are sensitive to the presence of topological defects.

The paper is structured as follows. In the next Section, we will
present our models, parametrisations and observables of interest. In
Section~\ref{sec:GW} we set out our numerical procedure and
definitions to compute the gravitational wave spectrum. In
Section~\ref{sec:results} we consider basic numerical observables,
including components of the energy-momentum tensor, as well as the
total energy generated in gravitational waves. We also present the
full spectrum of gravitational waves, and analyse new features and
contrast them with a scalar-only theory. We conclude in
Section~\ref{sec:discussion}, where we also provide the extrapolation
of the spectrum to the present day, and consider the potential for
detection by LISA.

\section{The $\mathrm{SU}(2)$-scalar model}
\label{sec:model}

We consider a complex scalar doublet $\phi$ (four real components),
coupled to an $\mathrm{SU}(2)$ gauge field $A_\mu$. We have the
action:
\begin{equation}
\label{eq:action}
S_{4+G} = -\int d^4 x \left\{\frac{1}{4g^2}
F^{a,\mu\nu} F_{a,\mu\nu} +
\left[\left(D_{\mu}\phi\right)^{\dagger}D^{\mu}\phi +\mu_{\rm eff}^2(t)\phi^\dagger\phi+
  \lambda(\phi^{\dag}\phi)^2\right] \right\},
\end{equation}
where
\begin{eqnarray}
F^a_{\mu\nu} = \partial_{\mu}A_{\nu}^a - \partial_{\nu}A_{\mu}^a +
\epsilon^{abc}A_{\mu}^bA_{\nu}^c,\qquad
D_{\mu}\phi = \left(\partial_{\mu} -iA^a_{\mu}\tau^a
\right)\phi,
\end{eqnarray}
and $\mu_{\rm eff}^2(t)$ is given by Eq.~(\ref{eq:mu2}). The scalar
field components are
\begin{equation}
  \phi = \frac{1}{\sqrt{2}} \left( \begin{array}{c} \phi_2 + i \phi_3
    \\ \phi_0 + i \phi_1 \end{array} \right).
\end{equation}
In the
following, we will refer to the scalar as the Higgs field. Although we
may only identify it with the Standard Model Higgs in the case where
$\mu\simeq$ 100 GeV, for any energy scale the symmetry breaking
transition will lead to an analogue of the electroweak transition.  In line with this terminology, we speak
of the Higgs mass $m_H=\sqrt{2}\mu$, the Higgs expectation value
$v=\mu/\sqrt{\lambda}$ and the W-mass (for the gauge fields),
$m_W=gv/2$. We will take the Standard Model values from $m_H=125$ GeV,
$m_W=80.2$ GeV, giving $\lambda\simeq 0.13$ and $g\simeq 0.65$. These
we will keep fixed while varying $\mu^2$ (and as a result, $v$). We
will also briefly discuss varying $\lambda$ (we consider
$\lambda=0.001$ and $\lambda=0.01$), leaving the other parameters
fixed. The lattice spacing will be fixed by $a\mu=0.17$, with a
lattice size of $N^3=384^3$ sites.

From the classical action (\ref{eq:action}), we can by variation with
respect to $\phi$ and $A_\mu$ derive the classical equations of
motion. These are explicit, coupled, non-linear, partial differential
equations that can be solved by discretisation on a spatial cubic
grid, and then evolved in time. We follow the standard procedure to do
this (see, for instance \cite{Enqvist:2015sua}).

In a cosmological context, we should in principle include the
expansion of the Universe through appropriately integrating the
Friedmann-Robertson-Walker metric in the action. However, we know that
the timescales involved are of the order of $10^2\mu^{-1}$, the total
energy density is $V_0=\mu^4/(4\lambda)$ and so we can ignore Hubble
expansion as long as
\begin{eqnarray}
1\gg\frac{100H}{\mu} =100\sqrt{\frac{1}{12\lambda}}\frac{\mu}{M_{\rm p}}.
\end{eqnarray} 
Energy scales $\mu \lesssim 10^{13}$ GeV satisfy this bound for our
choices of $\lambda$.

\subsection{Reduced models}
\label{sec:reduced}

We are mainly investigating non-Abelian gauge-Higgs systems, but for
comparison, we consider three other models, one with only a complex
scalar field, one with a complex scalar field coupled to a U(1) gauge
field and one with only a complex doublet\footnote{Simulations with a
  single-component real field produce domain walls stretching through the
  lattice. This is interesting, but not relevant for us here.},
(essentially an $\mathrm{O}(4)$-model). The actions are correspondingly for the
complex singlet (denoted by the label ``2'')
\begin{eqnarray}
S_2 = -\int d^4 x\, \left[(\partial_{\mu}\phi)^*\partial^\mu\phi +\mu_{\rm eff}^2(t)\phi^*\phi+
  \lambda(\phi^*\phi)^2\right],
\end{eqnarray}
for the U(1)-complex singlet
\begin{equation}
S_{2+G} = -\int d^4 x\, \left\{\frac{1}{4e^2}
F^{\mu\nu} F_{\mu\nu} +
\left[\left(D_{\mu}\phi\right)^{*}D^{\mu}\phi +\mu_{\rm eff}^2(t)\phi^*\phi+
  \lambda(\phi^*\phi)^2\right] \right\},
\end{equation}
where
\begin{eqnarray}
F_{\mu\nu} = \partial_{\mu}A_{\nu} - \partial_{\nu}A_{\mu},\qquad
D_{\mu}\phi = \left(\partial_{\mu} -iA_{\mu}
\right)\phi,
\end{eqnarray}
and for the complex doublet (denoted by the label ``4''),
\begin{eqnarray}
S_4 = -\int d^4 x\, 
\left[\left(\partial_{\mu}\phi\right)^{\dagger}\partial^{\mu}\phi +\mu_{\rm eff}^2(t)\phi^\dagger\phi+
  \lambda(\phi^{\dag}\phi)^2\right].
 \end{eqnarray}
Although in the following we describe the implementation and
observables for the $\mathrm{SU}(2)$-Higgs model (denoted by the label
``4+G''), these all apply with trivial adaptations to the 2, 2+G and 4 case. 

We note that with these standard conventions, the Higgs mass is always $m_H=\sqrt{2\lambda} v = \sqrt{2}\mu$, which we will keep the same for all models when comparing. On the other hand, whereas in the SU(2)-Higgs model the gauge field mass is $m_W=\frac{1}{2}gv$, in the U(1)-Higgs model, it is $m_W=e v$. When comparing, we will match the tree-level masses, and so take $e =g/2$.

\subsection{Initial conditions}
\label{sec:initcond}

We will model the cold spinodal transition by assuming that the
initial state is the vacuum in the potential
\begin{eqnarray}
V_{\rm in}(\phi)=\mu^2\phi^\dagger\phi,
\end{eqnarray}
so that the (free) modes of the Higgs field obey
\begin{equation}
\left< \phi_a(k)\phi_b(k)^\dagger \right> = \frac{1}{2\sqrt{\mu^2 + k^2}} \delta_{ab},\qquad 
\left< \pi_a(k)\pi_b(k)^\dagger \right> = \frac{1}{2}\sqrt{\mu^2 + k^2}\delta_{ab}.
\end{equation}
with $a$ denoting the four (two) real scalar degrees of freedom. We
only initialise unstable modes with $|{\bf k}|<\mu$. These are the
ones that grow large and subsequently validate the use of classical
dynamics rather than full quantum dynamics. Careful discussions of
this point can be found in
Refs.~\cite{GarciaBellido:2002aj,Smit:2002yg,Arrizabalaga:2004iw}. Consistent
with this way of thinking, all the gauge fields will be put to zero
initially $A_\mu=0$. Throughout, we will evolve the equations in
temporal gauge $A_0=0$. We also initialise the gauge field conjugate
momenta $E_i$ to zero. There is therefore an insignificant residual
per-site violation of the Gauss law of relative order $10^{-8}$.

These initial conditions are closely related to those of
Ref.~\cite{Dufaux:2010cf}. However, we note that although one may
classically scale out the vacuum expectation value $v$ from the
classical equations of motion for the scalar, one cannot in the same
way rescale the quantum initial conditions. These are the same for any
choice of $v$, provided $\mu$ is fixed. On the lattice they are
determined by the choice of lattice scale. Hence, one may not
trivially scale the results from one value of $v$ to another, since
the initial condition is then de facto different (although the
resulting error is probably small). On the other hand, one may compute
dimensionless ratios at a given lattice spacing (defined for instance
in units of the mass $a\mu$), and then rescale trivially in $\mu$. We
will do this below.

\subsection{Observables and Energy-Momentum tensor}
\label{sec:EM}

The energy-momentum tensor of the theory follows from variation with
respect to the metric. For the scalar field, we have for the energy
density
\begin{eqnarray}
\rho_\phi = (\partial_t\phi)^\dagger\partial_t\phi + (D_i\phi)^\dagger D_i\phi+\mu_{\rm eff}^2(t)\phi^\dagger\phi+\lambda(\phi^\dagger\phi)^2+ V_0.
\end{eqnarray}
We may further subdivide this contribution into a kinetic (first
term), gradient (second term) and potential part (the
rest). Initially, the bulk of the energy is in $V_0$. An important
consequence of our quench mechanism is that total energy is not
conserved, because $\mu_{\rm eff}^2(t)$ has an explicit
time-dependence. We have that
\begin{eqnarray}
\label{eq:Eloss}
\frac{dE}{dt}= \frac{d\mu_{\rm eff}^2(t)}{dt}\int d^3 {\bf x} \,\phi^\dagger \phi({\bf x},t).
\end{eqnarray}
For the largest quench times presented here, this leads to a sizeable
depletion of energy of up to 75\%, or reduction of the final
temperature of 50 \%. This is the price we pay for simplifying the
system by ignoring the specific and model-dependent dynamics of the
trigger (inflaton) field.

It turns out \cite{Mou:2017xbo} that such a quench, where energy is
initially taken out of the Higgs field, corresponds to a particular
parameter subspace of the model given in Eq.~(\ref{eq:portal}). At
later times, energy is reintroduced as the system equipartitions and
equilibrates. The time-scale for this to complete is much longer than
the time-scales considered here. The late-time dynamics can also
generate gravitational waves (see for instance \cite{Dufaux:2010cf}).

Gravitational waves are sourced by the off-diagonal spatial components
$T^\phi_{ij}$, given by
\begin{equation}
T^\phi_{ij} = (D_i \phi)^\dag (D_j \phi) + (D_i \phi) (D_j\phi)^\dag = 2 \mathrm{Re} \left[(D_i \phi)^\dag (D_j \phi)\right].
\end{equation}
For the gauge field, we have 
\begin{eqnarray}
T_{\mu\nu}^A= \eta_{\mu\nu}\mathcal{L} -\frac{1}{g^2} \eta^{\alpha\beta}F_{\mu\alpha}^aF_{\nu\beta}^a.
\end{eqnarray}
The energy density is
\begin{eqnarray}
\rho_A=\frac{1}{g^2} \left(F^{a}_{0\alpha}F^{a}_{0\alpha}+\frac{1}{4}F^{a,\mu\nu}F_{a,\mu\nu}\right)=\frac{1}{2}(E_i^2+B_i^2).
\end{eqnarray}
The off-diagonal spatial components entering the gravitational wave
computation are then
\begin{eqnarray}
T_{ij}^A=-\frac{1}{g^2} F_{i}^{\beta,a}F_{j\beta}^a.
\end{eqnarray}
The complex singlet and doublet models for comparison have no gauge
field contribution, and the scalar energy-momentum expressions include
normal instead of covariant derivatives, with two and four real
fields, respectively.

\section{Gravitational wave production}
\label{sec:GW}

Given the background scalar-gauge field theory simulation, from which
we extract the energy-momentum tensor as described above, we can
compute the gravitational wave spectrum as described in the
following. For a more detailed exposition we refer to
Ref.~\cite{Hindmarsh:2015qta}, and references therein.  At each step
of the simulation we compute the parts of the stress energy tensor
that source metric perturbations, namely $T^A_{ij}$ for the gauge
field and $T^\phi_{ij}$ for the Higgs field. We can then numerically
explicitly solve the wave equation for the metric perturbation
$u_{ij}$~\cite{GarciaBellido:2007af},
\begin{equation}
\label{eq:ueq}
{\ddot u}_{ij} - \nabla^2 u_{ij} = 16 \pi G (T^\phi_{ij} + T^{A}_{ij}).
\end{equation}
Going to momentum space
\begin{equation}
u_{ij}({\bf k}) = \int d^3 {\bf x}\, u_{ij}(\mathbf{x}) e^{-i \mathbf{k}\cdot \mathbf{x}},
\end{equation}
we can then project out the propagating, transverse-traceless degrees
of freedom
\begin{equation}
h_{ij} (t,\mathbf{k}) = \lambda_{ij,lm}(\hat{\mathbf{k}}) u_{lm}(t,\mathbf{k}),
\end{equation}
where 
\begin{eqnarray}
\lambda_{ij,lm}({\bf k}) = P_{ik}({\bf k})P_{jl}({\bf k})-\frac{1}{2}P_{ij}({\bf k})P_{kl}({\bf k}),\qquad P_{ij}({\bf k})= \delta_{ij}-\frac{k_ik_j}{|{\bf k}|^2}.
\end{eqnarray}
We can then construct the total energy density in the gravitational waves
\begin{equation}
\label{eq:totalpower}
\rho_\text{GW} =
\frac{1}{32 \pi G V} \int \frac{d^3 {\bf k}}{(2\pi)^3} \langle \dot{h}^{ij}(\mathbf{k}) \dot{h}^{ij}(-\mathbf{k})\rangle,
\end{equation}
where the average is to be taken over the full quantum state, or in
practice an ensemble of realisations of the field theory initial
conditions.  Our ensembles of realisations are quite small, $\mathcal{O}(10)$ field configurations, since the convergence of the average turns out to be quite fast.

We may also define the spectrum of gravitational waves,
\begin{equation}
\label{eq:spectrum}
\frac{d \rho_\text{GW}}{d\ln k} = \frac{1}{32\pi G V} \frac{k^3}{(2\pi)^3}
\int d\Omega\, \langle \dot{h}^{ij}(\mathbf{k}) \dot{h}^{ij}(-\mathbf{k})\rangle,
\end{equation}
where the integral is now only over solid angle. The spectrum is then
a function of the length of ${\bf k}$ only. Note the contraction of
labels $ij$ in Eqs.~(\ref{eq:totalpower}) and~(\ref{eq:spectrum}).

We solve Eq.~(\ref{eq:ueq}) in parallel as the simulation of the
(gauge-)Higgs system is performed. The source terms are
time-dependent, and are in effect integrated over time to produce the
final gravitational wave spectrum, and the total energy density in
gravitational waves. This energy density is numerically completely
negligible relative to the total energy density of the field theory
system, and it makes little sense feeding the created gravitational
waves back into the field theory simulation. Hence the gravitational
waves are computed in the background of the (gauge-)Higgs system with
no back-reaction. Ultimately, it is the quantity given by
Eq.~(\ref{eq:spectrum}) that may be inferred from observations,
suitably transported from the end of inflation to the present
time. This is discussed further in Section~\ref{sec:discussion}.

\section{Results}
\label{sec:results}

\subsection{Energy distribution and total gravitational wave power}
\label{sec:resultsA}

\begin{figure}
\includegraphics[width=0.5\textwidth]{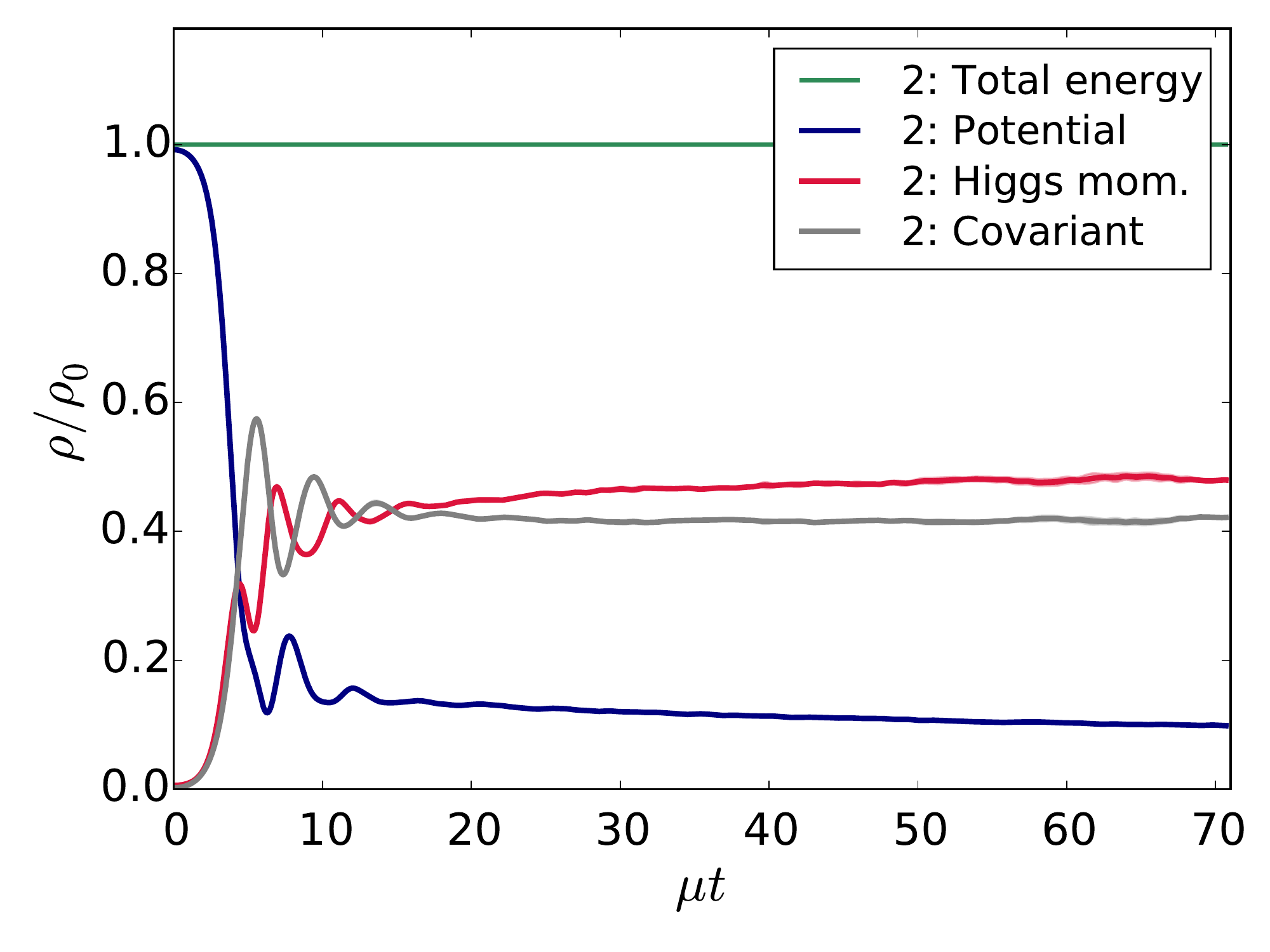}
\includegraphics[width=0.5\textwidth]{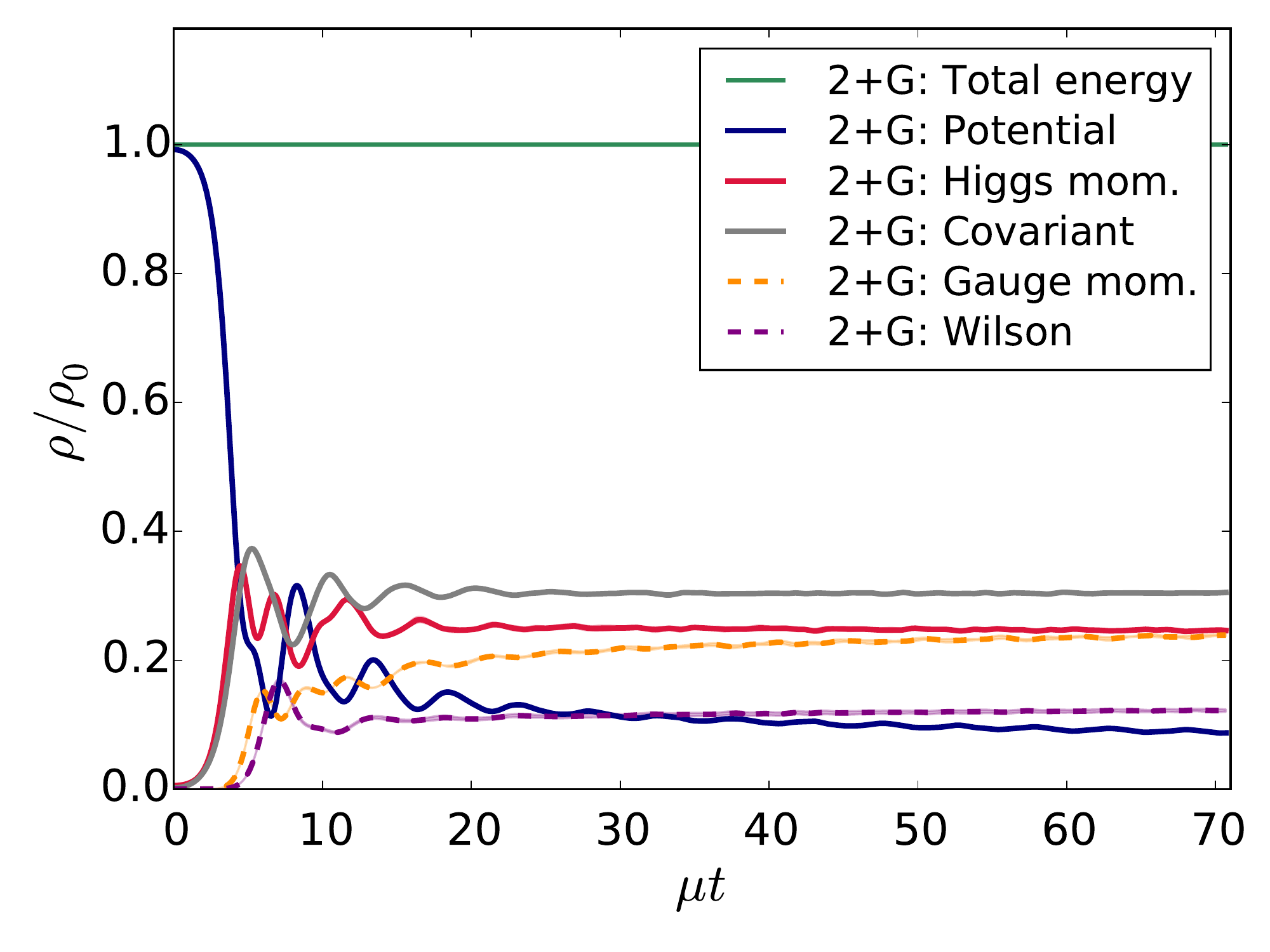}
\includegraphics[width=0.5\textwidth]{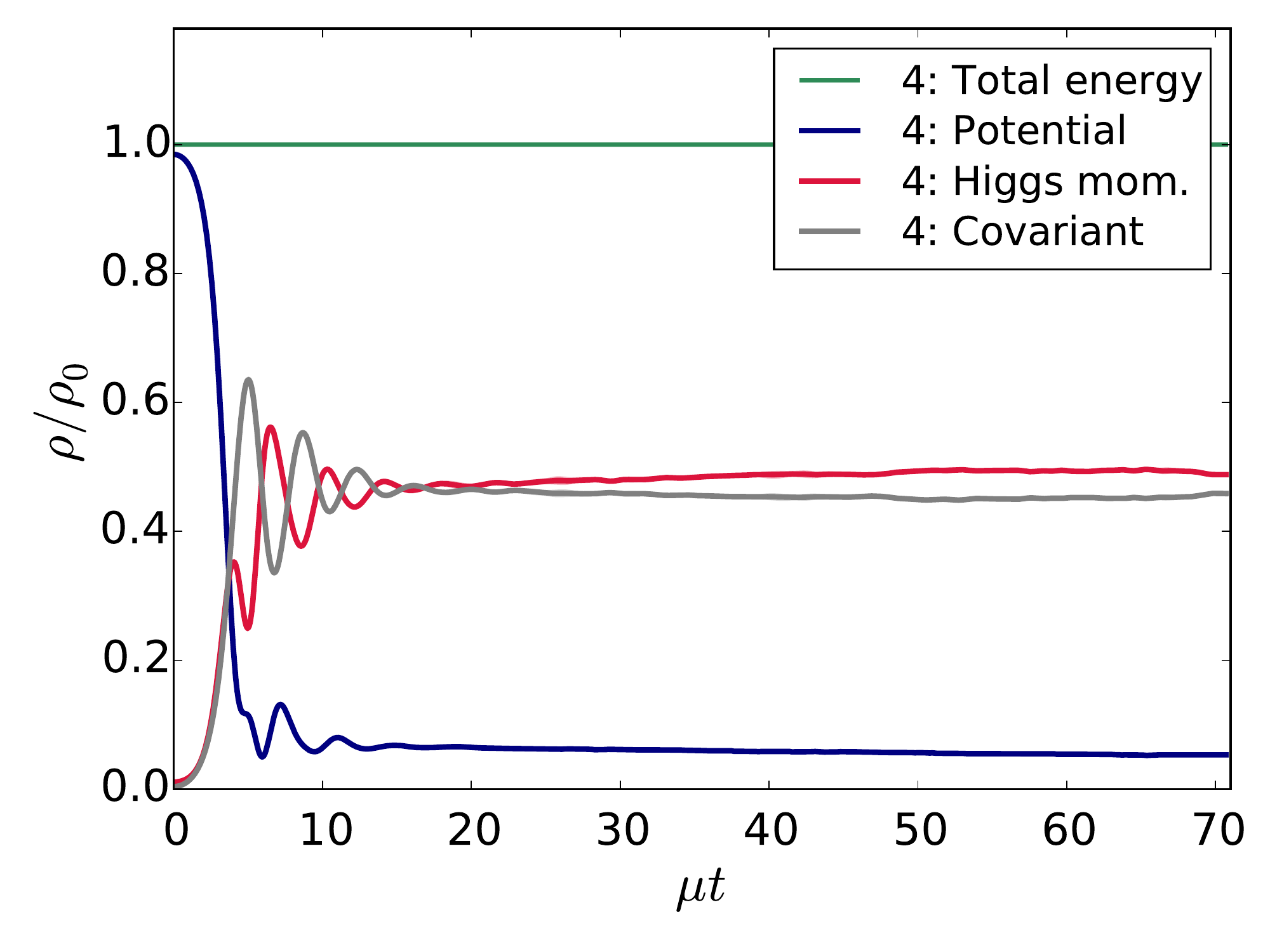}
\includegraphics[width=0.5\textwidth]{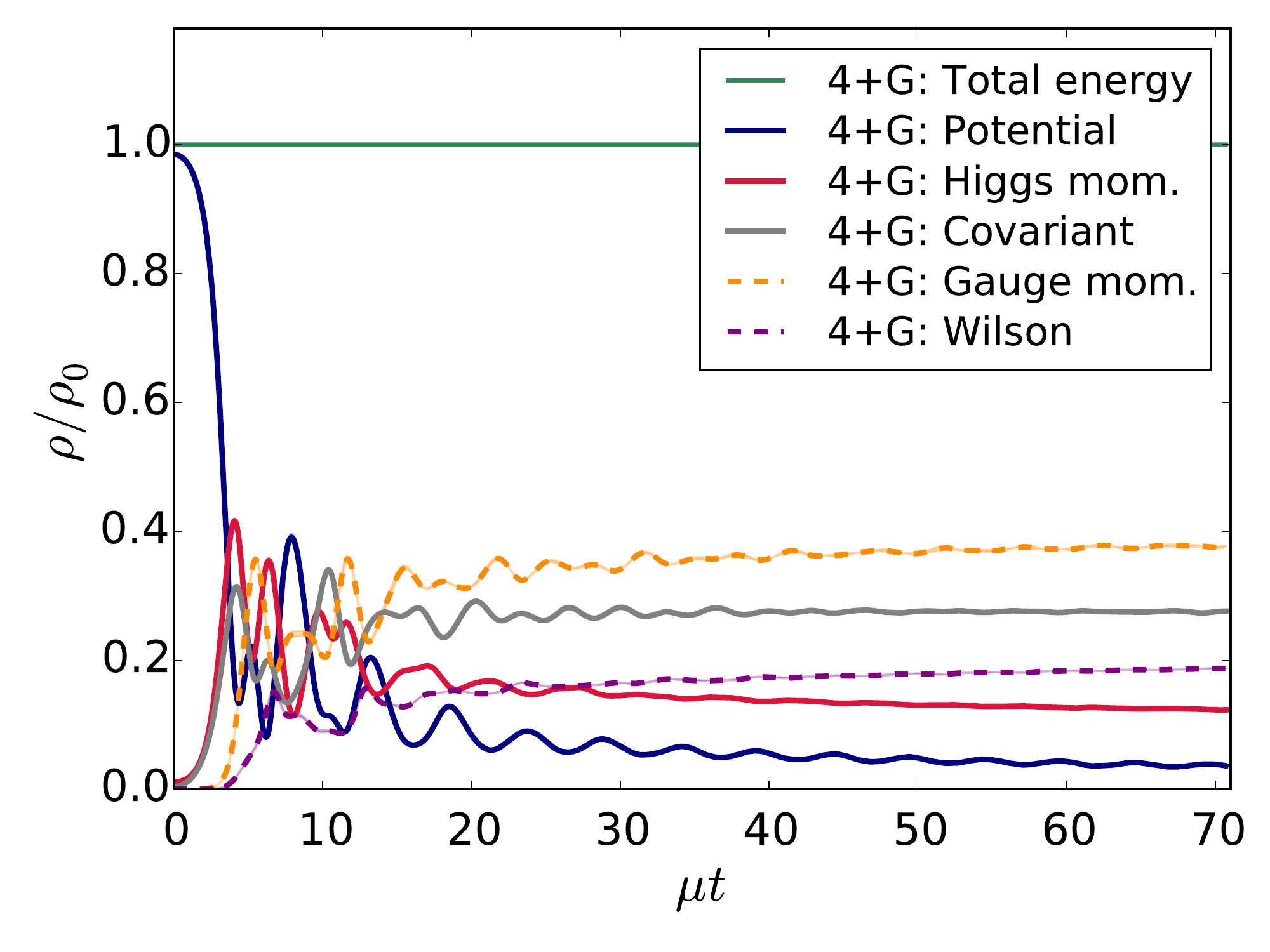}
\caption{The energy components of the complex Higgs  (top left), U(1)-Higgs (top right), doublet Higgs (bottom left) and SU(2)-Higgs (bottom right) systems. Quench time is $\mu \tau_q=0$. }
\label{fig:Excomponents}
\end{figure}
Gravitational waves are sourced by the off-diagonal components of the
energy-momentum tensor, but it is instructive to consider the energy
density and its components, to track where the energy goes. Initially,
all the energy, except for small fluctuations in the Higgs field, is
in the Higgs potential $V_0$. In a tachyonic transition, the low
momentum modes $|{\bf k}|<\mu$ grow exponentially, picking up kinetic
energy and gradient energy, while losing potential energy. With gauge
fields coupled to $\phi$, these will also grow exponentially, and reheat
simultaneously \cite{Skullerud:2003ki}.

In Fig.~\ref{fig:Excomponents}, we show the various energy components
for a simulation at one particular quench time $\mu\tau_q=0$. In the
top left panel, we show simulations with a complex scalar, top right
when adding a U(1) gauge field. On the bottom left is the doublet
scalar, bottom right when adding and SU(2) gauge field.  We see that
80\% of the potential energy is transferred to kinetic and gradient
energy within $5 \mu^{-1}$. There is some quantitative difference
between the singlet and doublet case, but qualitatively they are very
similar. The transition is over after $\mu t\simeq 15-20$. Note that
for this case of zero quench time, no energy is lost because of the
quenching process (\ref{eq:Eloss}).

On the right-hand panels of the figure, we see that including the
gauge field changes the situation. Although the potential energy is
released very quickly as for the pure-scalar cases, this transfer only
completes somewhat later. The Higgs field oscillates more and for
longer and has a smaller fraction of the total energy. Some of this
energy is instead transferred to the gauge field, which takes longer
to be excited and settle, lasting until $\mu t\simeq 20-25$. There
also seems to be a qualitative difference between $\mathrm{U}(1)$ and
$\mathrm{SU}(2)$. For the Abelian gauge field, there is less energy in
the gauge field than in the Higgs field (orange/purple compared to
grey, red and blue). For SU(2), it is the other way around.

A detailed analysis of this preheating process in the SU(2)-Higgs
model can be found in
\cite{Skullerud:2003ki,GarciaBellido:2003wd}. The main features are
that there is a short, violent roll-off period, followed by kinetic
equilibration (and equipartition) as the self-interactions kick in,
with a time-scale of a few hundreds in mass units. A
Bose-Einstein-like particle spectrum is created with a sizeable
effective chemical potential which disappears on a time-scale of a few
thousands in mass units, through chemical equilibration.

\begin{figure}
  \centering
  \includegraphics[width=0.5\textwidth]{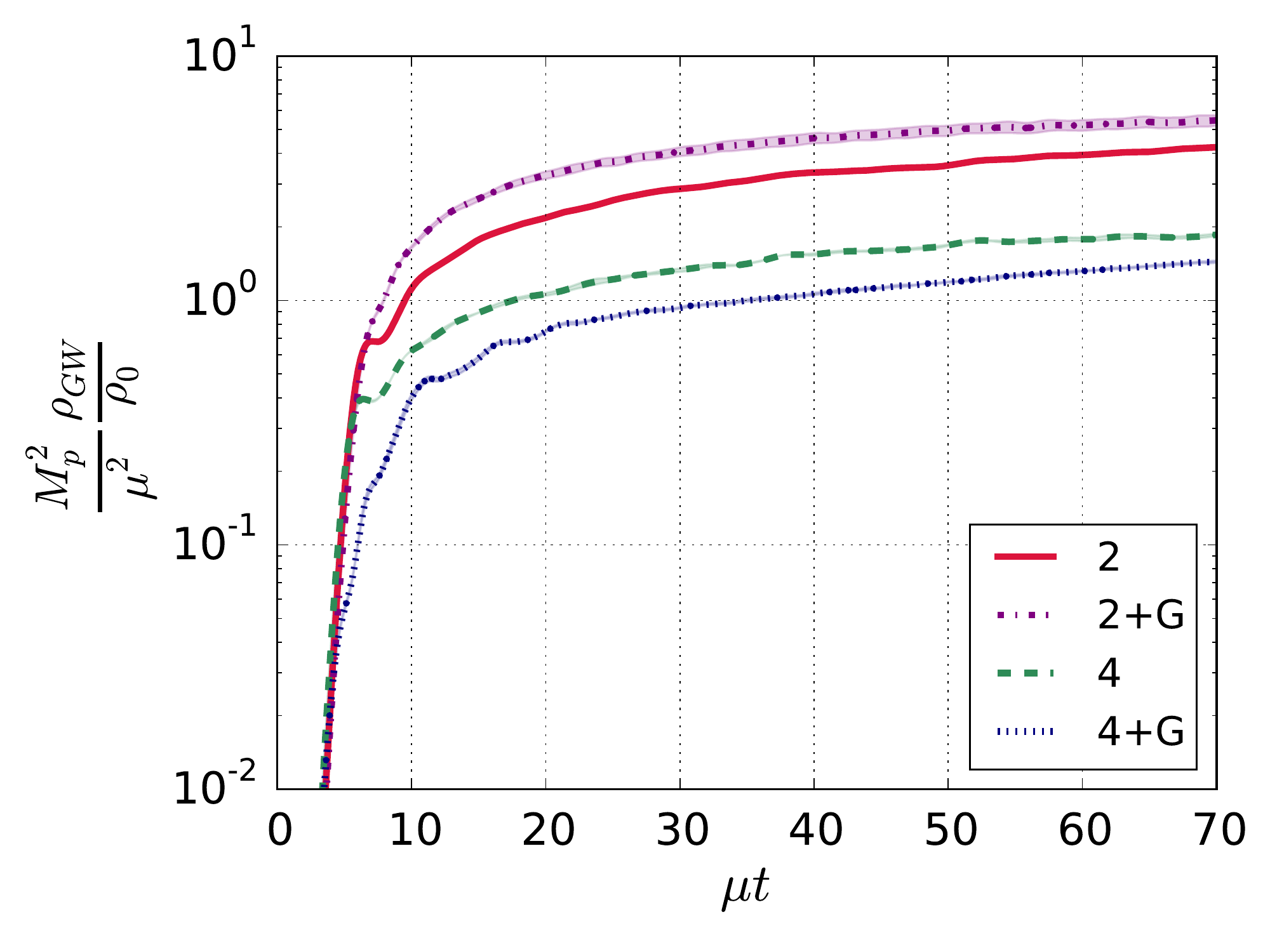}
  \caption{The total gravitational wave energy density for the four
    models under consideration, with $\mu \tau_q = 0$.}
  \label{fig:GWTheories}
\end{figure}
We start by comparing the total energy density in gravitational waves
for the four models under consideration in
Figure~\ref{fig:GWTheories}. It is interesting to note that, whereas
the addition of a U(1) gauge field increases the total gravitational
wave energy density, the SU(2) gauge field suppresses it. This is an
issue that we shall return to later.

\begin{figure}
\includegraphics[width=0.5\textwidth]{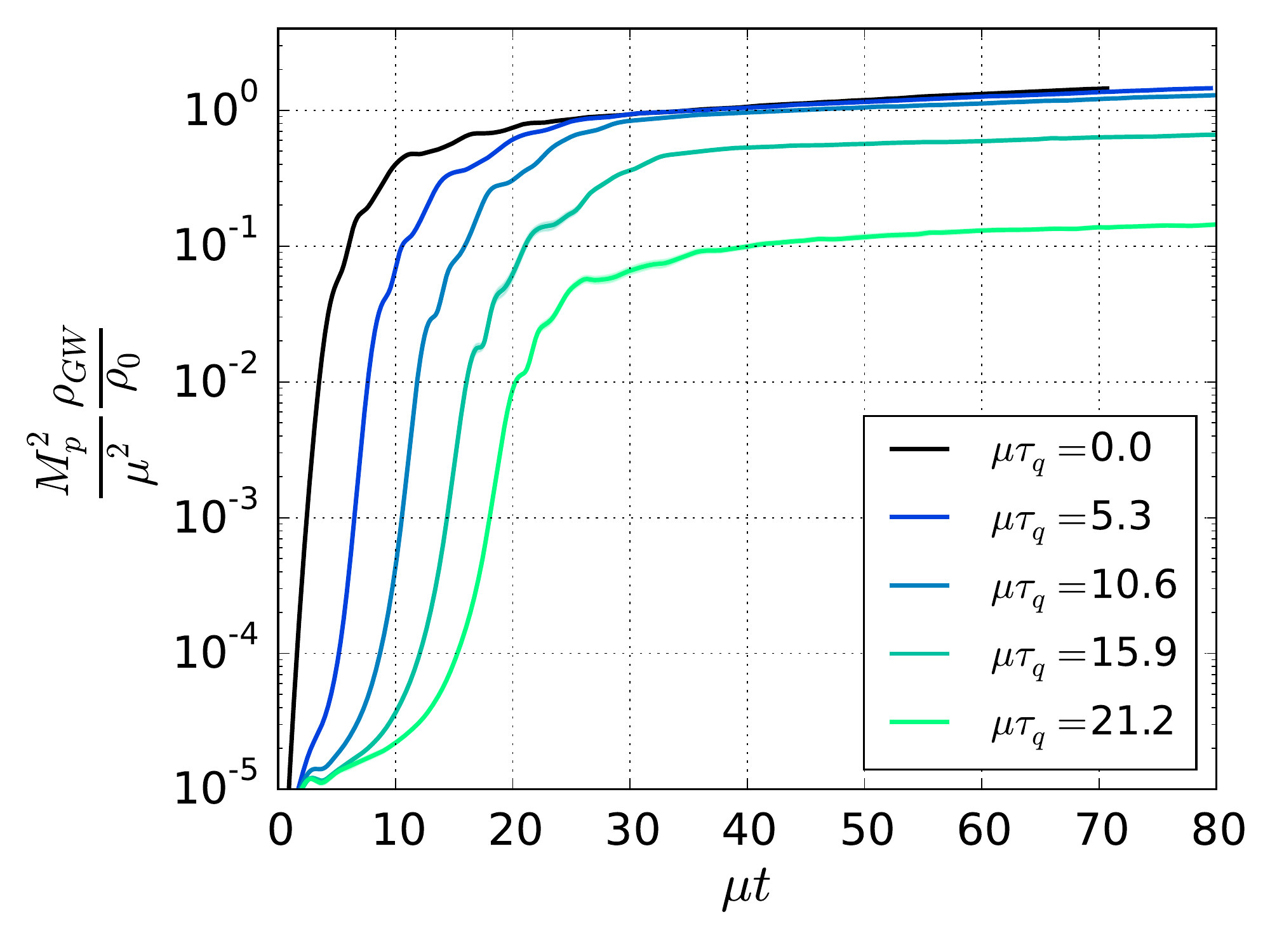}
\includegraphics[width=0.5\textwidth]{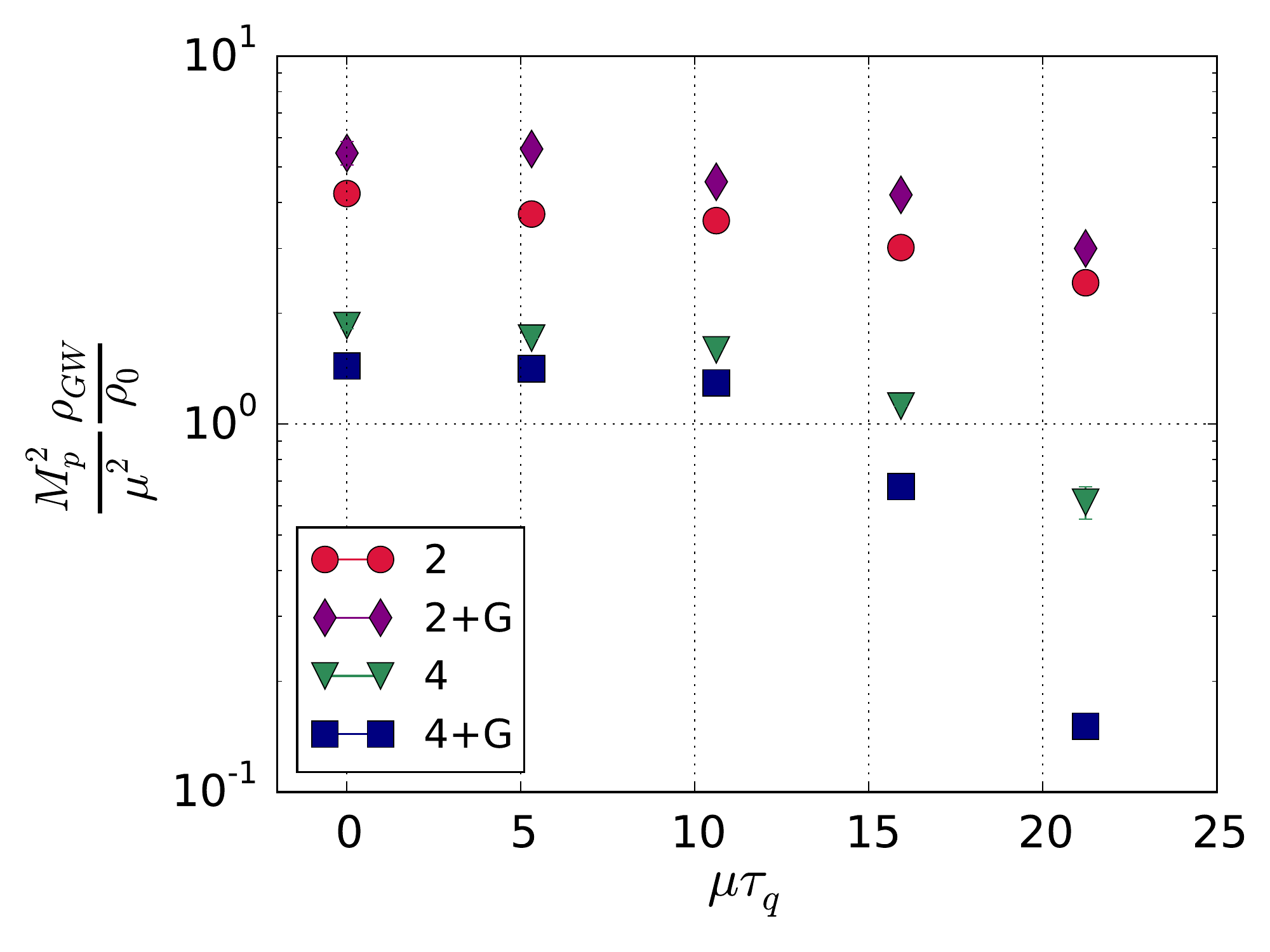}
\caption{The total gravitational wave energy density for the SU(2)-Higgs model for five quench times (left). And the final value, for all quench times and models (right).}
\label{fig:CompGWpower}
\end{figure}
In Fig.~\ref{fig:CompGWpower} (left), we show the total energy in
gravitational waves from gauge-scalar simulations for five different
quench times. The energy density is normalised to the initial energy
in the Higgs potential, and with a prefactor $\mu^2/M_{\rm p}^2$, with
$M_{\rm p}$ the Planck mass. For a given value of $\mu$, one should
therefore rescale the curves in the plot accordingly. As an example,
the Standard Model has $\mu\simeq 88 $ GeV, giving a prefactor of
$1.3\times 10^{-33}$.

We see that the violent transition causes the gravitational energy to
grow exponentially until a time $\mu t\simeq 10$ after the quench, and
that it continues to grow slowly afterwards.  We also see that the
final total power at first has little dependence on quench time, and
then decreases with quench time. This shows that when the time-scale
of the quench is below a certain cut-off, the time-scale of the
dynamics is the spinodal roll-off itself, rather than the quench time. 

In Fig.~\ref{fig:CompGWpower} (right) we show the total gravitational
wave energy for all four models, for all quench times. The total
energy is independent of quench time for $\mu \tau_q>10$, and then
starts decreasing for slower quenches. Remarkably, the GW production
increases when adding U(1) gauge fields to the complex scalar, whereas
it decreases when adding SU(2) gauge fields to the doublet scalar (and
more so for slower quenches). It seems that shifting energy into the
self-interacting non-Abelian gauge field has the effect of reducing GW
production.

\subsection{Gravitational wave spectrum}
\label{sec:resultsB}

\begin{figure}
\includegraphics[width=0.5\textwidth]{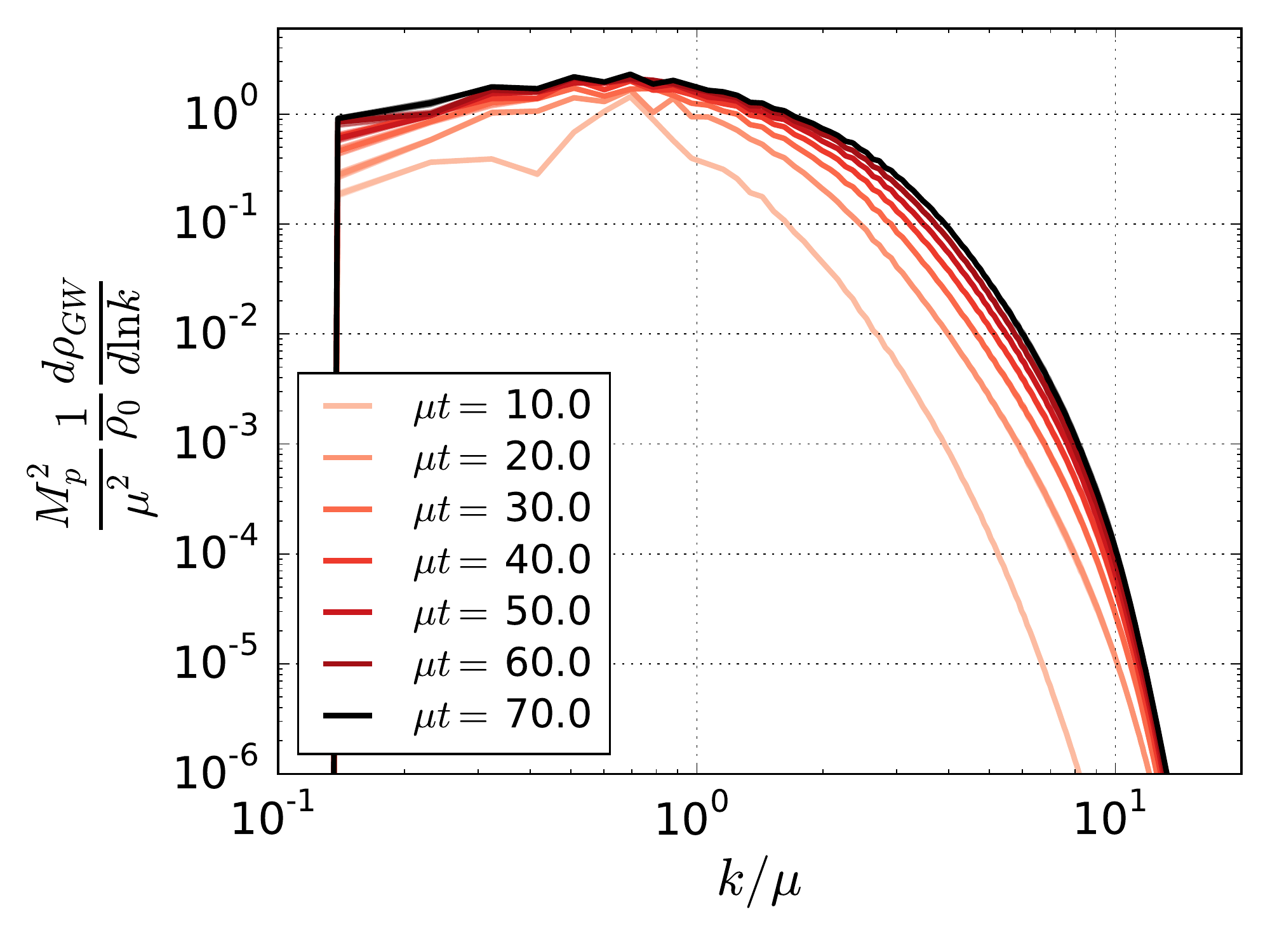}
\includegraphics[width=0.5\textwidth]{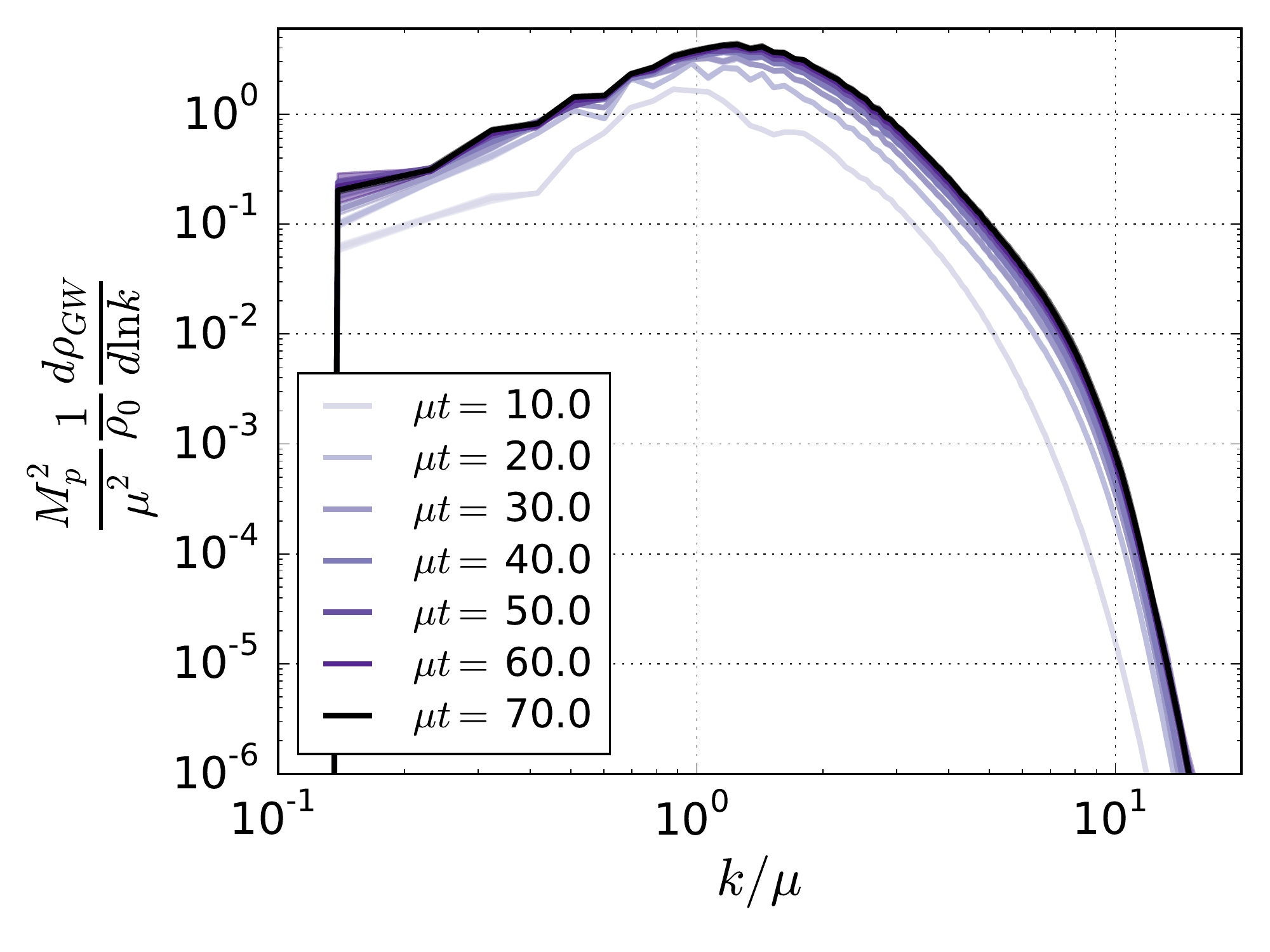}
\includegraphics[width=0.5\textwidth]{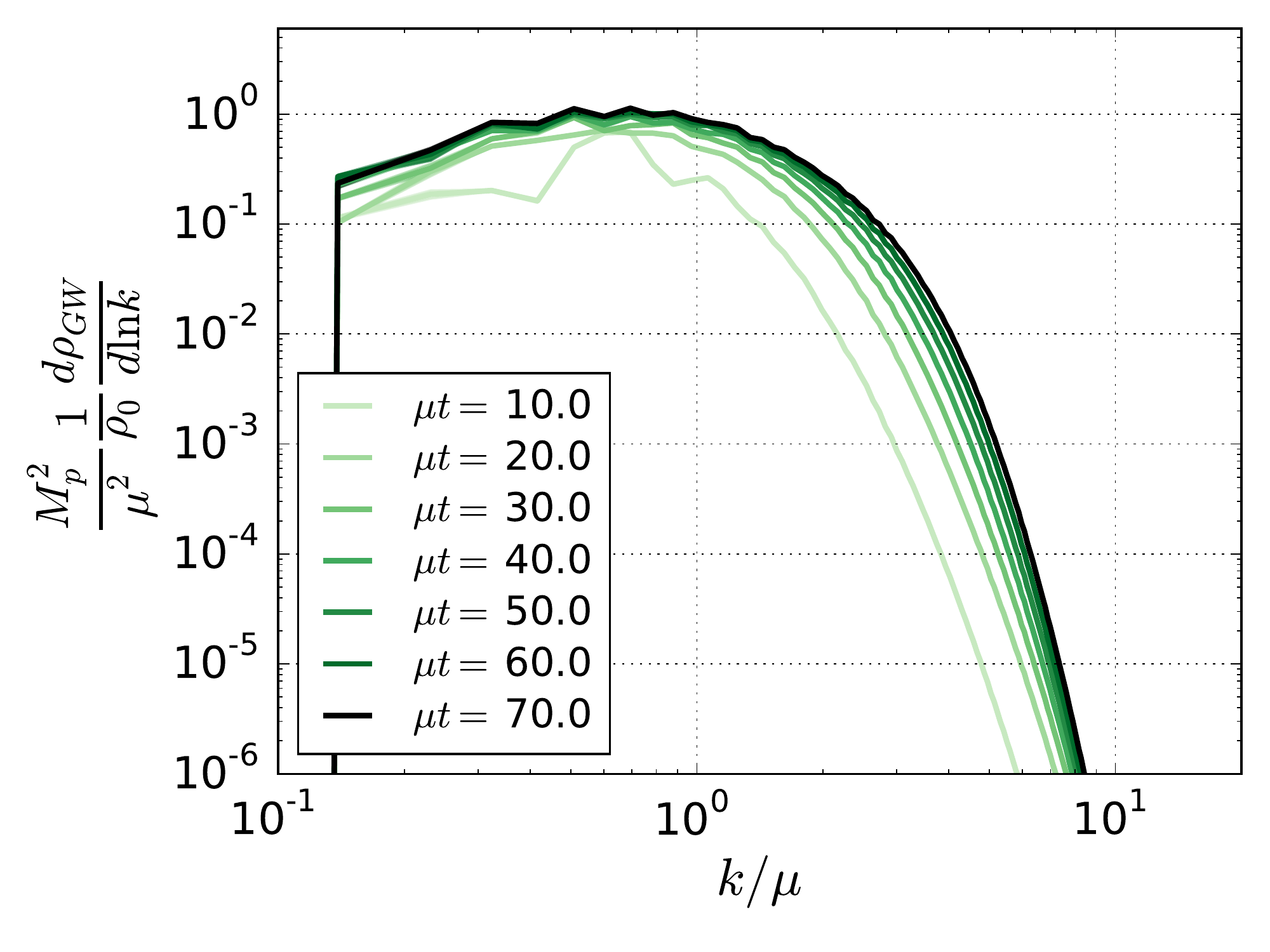}
\includegraphics[width=0.5\textwidth]{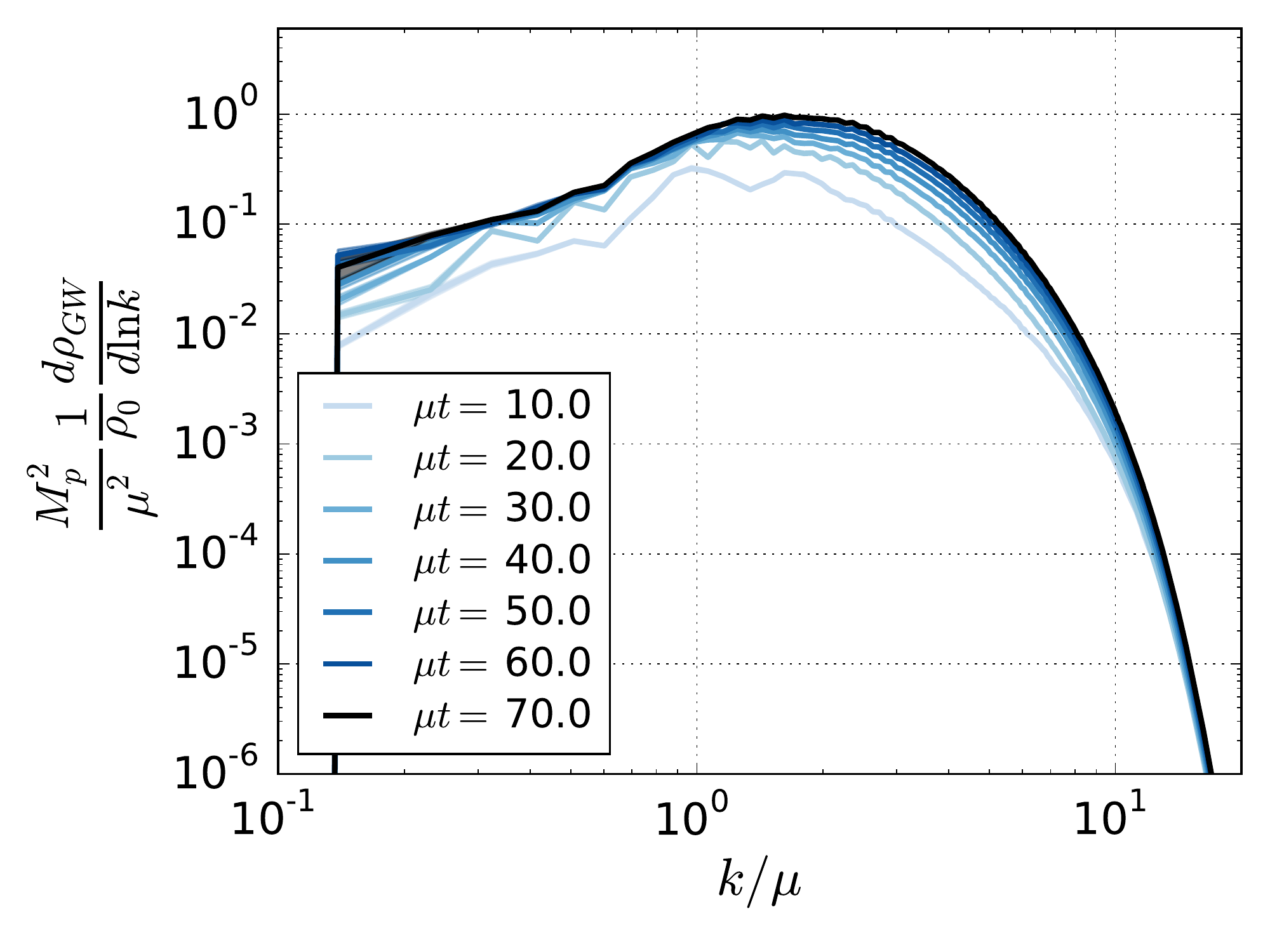}
\caption{The power spectrum for different times, for the complex Higgs  (top left), U(1)-Higgs (top right), doublet Higgs (bottom left) and SU(2)-Higgs (bottom right). Quench time is $\mu\tau_q=0$.}
\label{fig:Spect}
\end{figure}
Having found the total power in gravitational waves and its dependence
on quench time, we now proceed to study in detail the power spectrum
of gravitational waves produced by the transition. In
Fig.~\ref{fig:Spect}, we show the spectra for different simulation
times, fixing quench time $\mu\tau_q=0$, for our four cases: a complex
scalar (top left), when adding a U(1) gauge field (top right), for a
doublet scalar (bottom left) and when adding and SU(2) gauge field to
that (bottom right). We have again multiplied by a factor of $M_{\rm
  p}^2/\mu^2$, to be scaled back once a value for $\mu$ is chosen. The
spectrum grows and converges in shape and magnitude at time $\mu
t\simeq 60$.

For the scalar-only simulations, the spectrum is very similar, with a
peak around $k/\mu=0.7$ and an amplitude of about 1-2. The U(1)-Higgs
spectrum has a bit more power in the UV. When adding gauge fields, the
peak shifts to about $k/\mu=1.4-1.6$, with the U(1)-Higgs peak
becoming more pronounced. But whereas the U(1)-Higgs maximum is twice
its scalar-only counterpart, for the SU(2)-Higgs model the amplitude
does not change when adding gauge fields.

\begin{figure}
\includegraphics[width=0.5\textwidth]{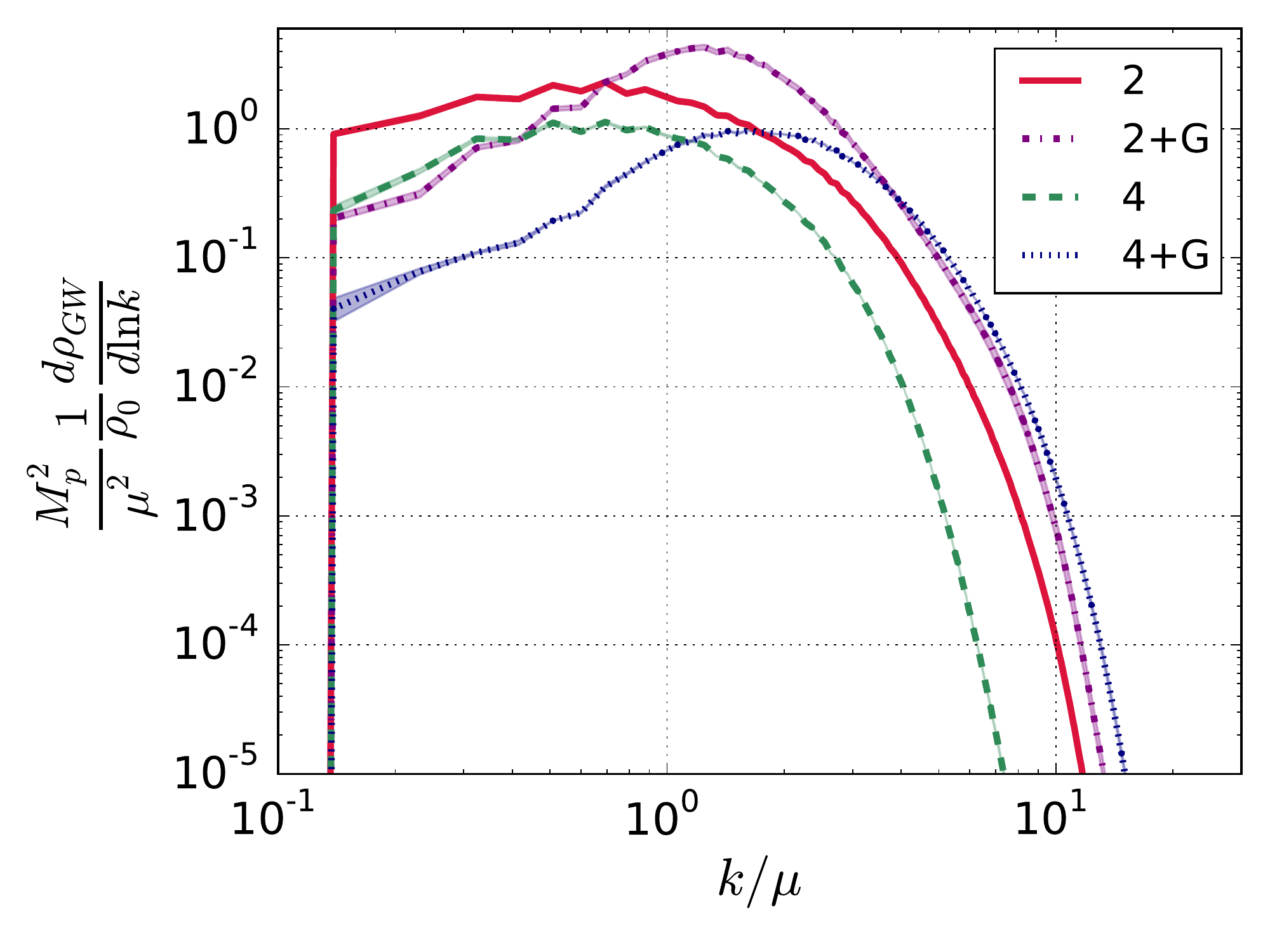}
\includegraphics[width=0.5\textwidth]{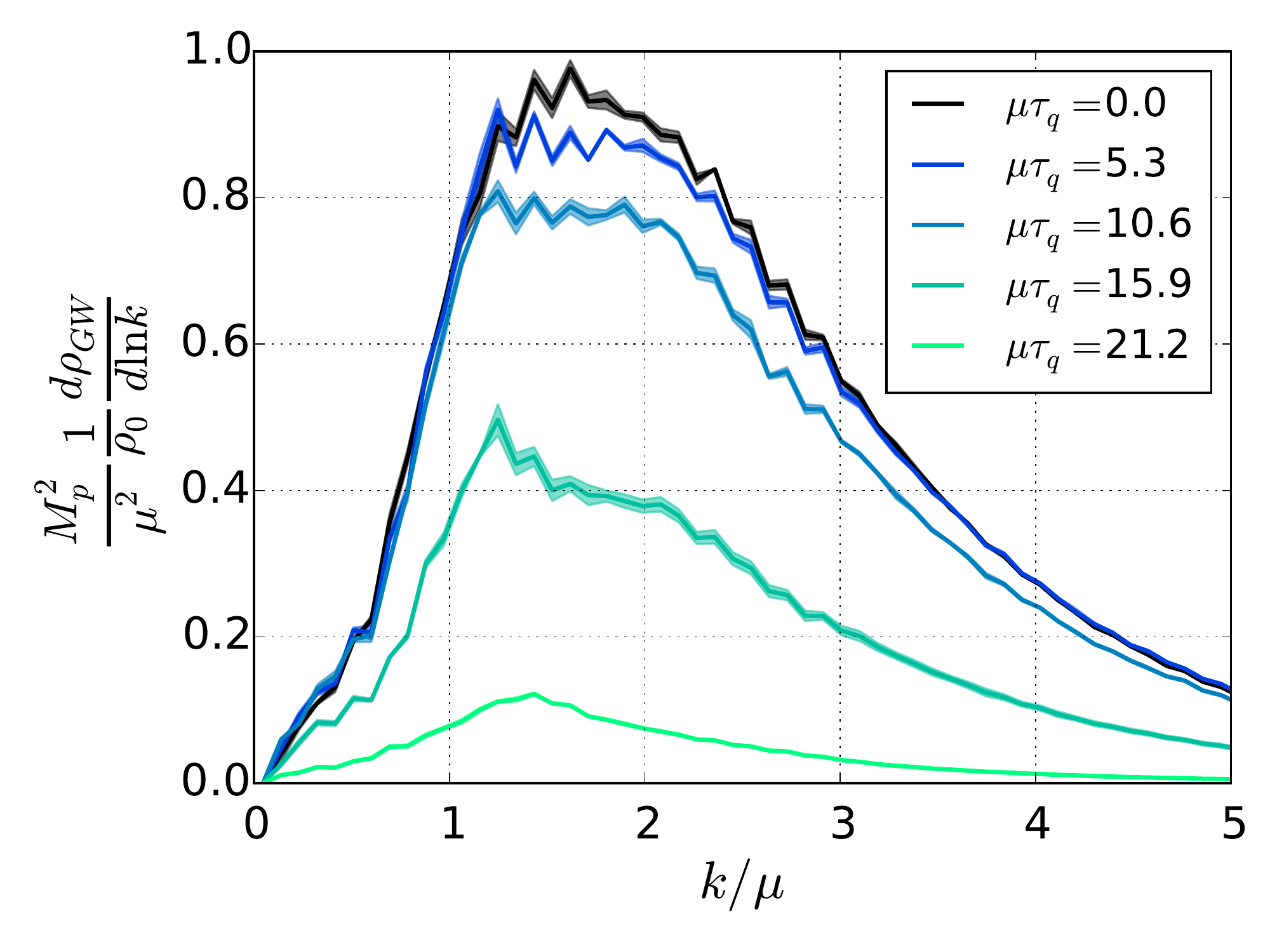}
\caption{The final spectrum all four models (left). And for the SU(2)-Higgs case, for different quench times (right)}
\label{fig:SpecComp}
\end{figure}
In Fig.~\ref{fig:SpecComp} (left) we show the spectra at the final
time of $\mu t=70$. We again see the shift in peak position and amplitude when adding gauge fields. As the peak moves to the right, more of the IR-part of the spectrum is revealed, exhibiting a power law dependence.

In the right-hand plot of the same figure, we then show the final
spectrum for the gauge-doublet case for different quench times (this
time on a linear scale).  We see that the peak value decreases
monotonically with longer quench time, although again the very quick
quenches cannot be resolved by the field dynamics. Note that since
gravitational wave production continues in principle indefinitely, we
have chosen to compare different quench time results at equal time
after the end of the quench, $\mu t_{\rm final}=\mu\tau_q+70$. The
slowest quench has a peak amplitude down by a factor of 8.

\begin{figure}
\includegraphics[width=0.5\textwidth]{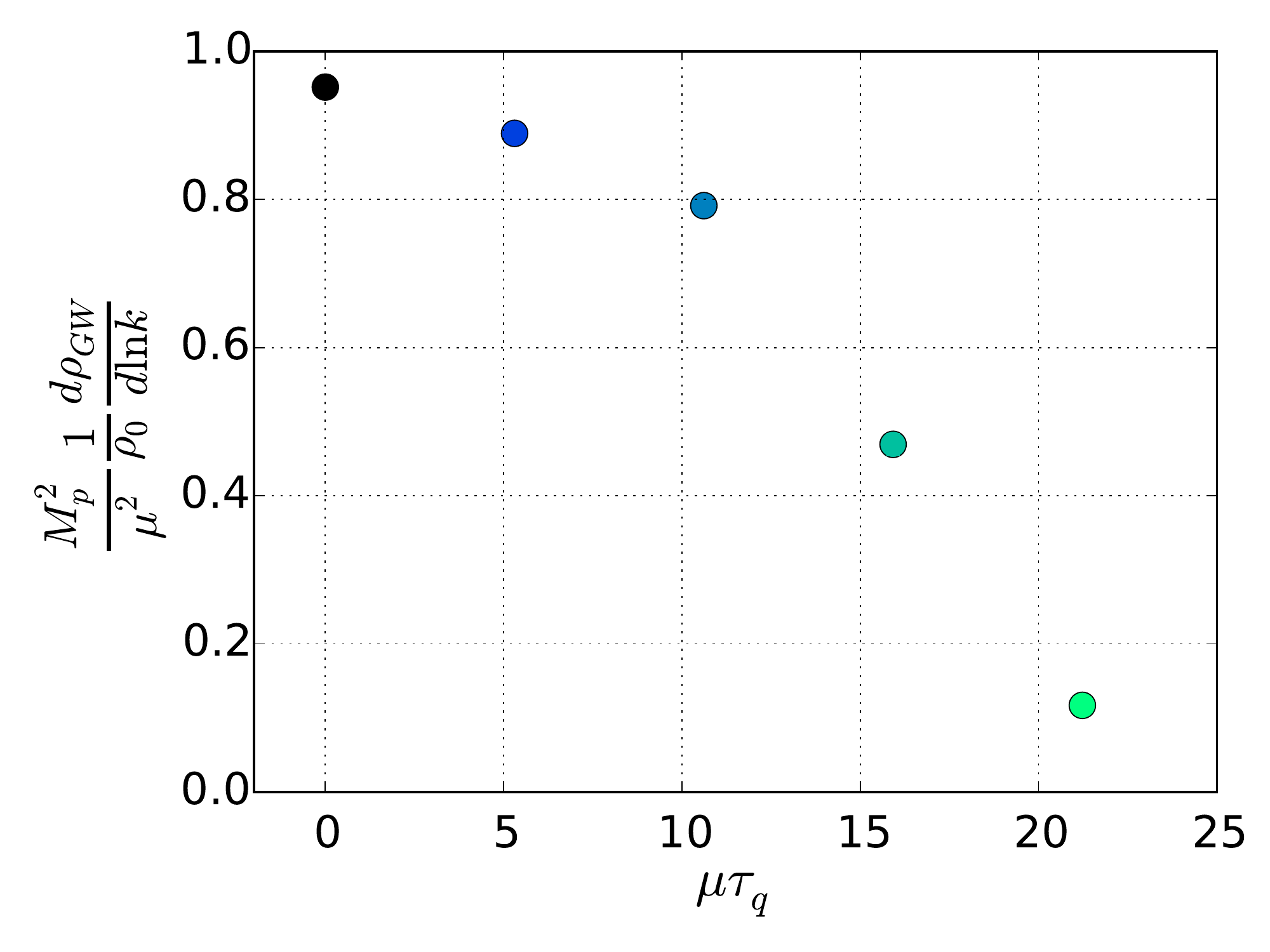}
\includegraphics[width=0.5\textwidth]{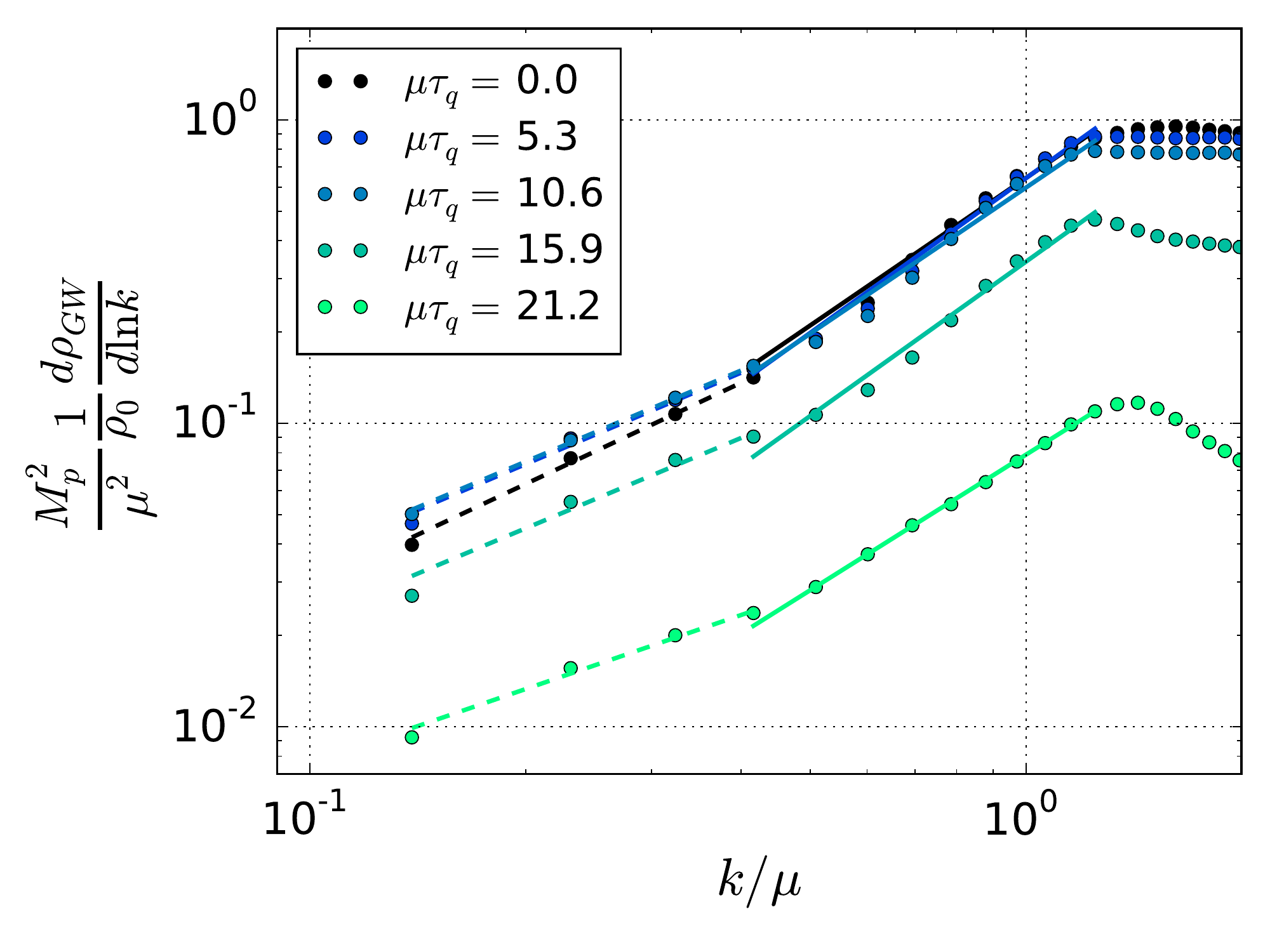}
\caption{The peak amplitude (left) and IR slope (right) for different quench times. SU(2)-Higgs model only. }
\label{fig:Features}
\end{figure}
Finally, we can attempt an analysis of the peak of the spectrum for
the gauge-doublet case, shown in Fig.~\ref{fig:Features}. The peak
position varies in the range $k/\mu = 1.4-1.6$ (not shown). The peak
amplitude (left-hand plot) shows a decreasing trend as a function of
quench time, reminiscent of the total energy density in gravitational
waves, Fig.~\ref{fig:CompGWpower}. One may also attempt a fit of the
IR slope(s) of the peak (right-hand plot) to find consistently a power
of $1.5-1.75$ for all quench times on the range
$\mu/2<k<\mu$. However, this power law is replaced by a shallower
$k-$dependence further in the IR, $k<\mu/2$. This far-IR power law is
in the range $0.8-1.1$ and generally close to unity. Further still, at
scales too large to study in our lattice simulations, this will give
way to the causal $k^3$ power law.

\subsection{Varying $\lambda$}
\label{sec:varlam}

So far, all our simulations have been done with $\lambda=0.13$, the
Standard Model value. The Standard Model is special in that the gauge
boson and Higgs masses are very similar
\begin{eqnarray}
\frac{m_H}{m_W}=\sqrt{\frac{8\lambda}{g^2}} \simeq 1.57.
\end{eqnarray}
This means that although there are two mass-scales in the problem (in
addition to quench time), we only see one peak in the spectrum of
gravitational waves.

\begin{figure}
\includegraphics[width=0.5\textwidth]{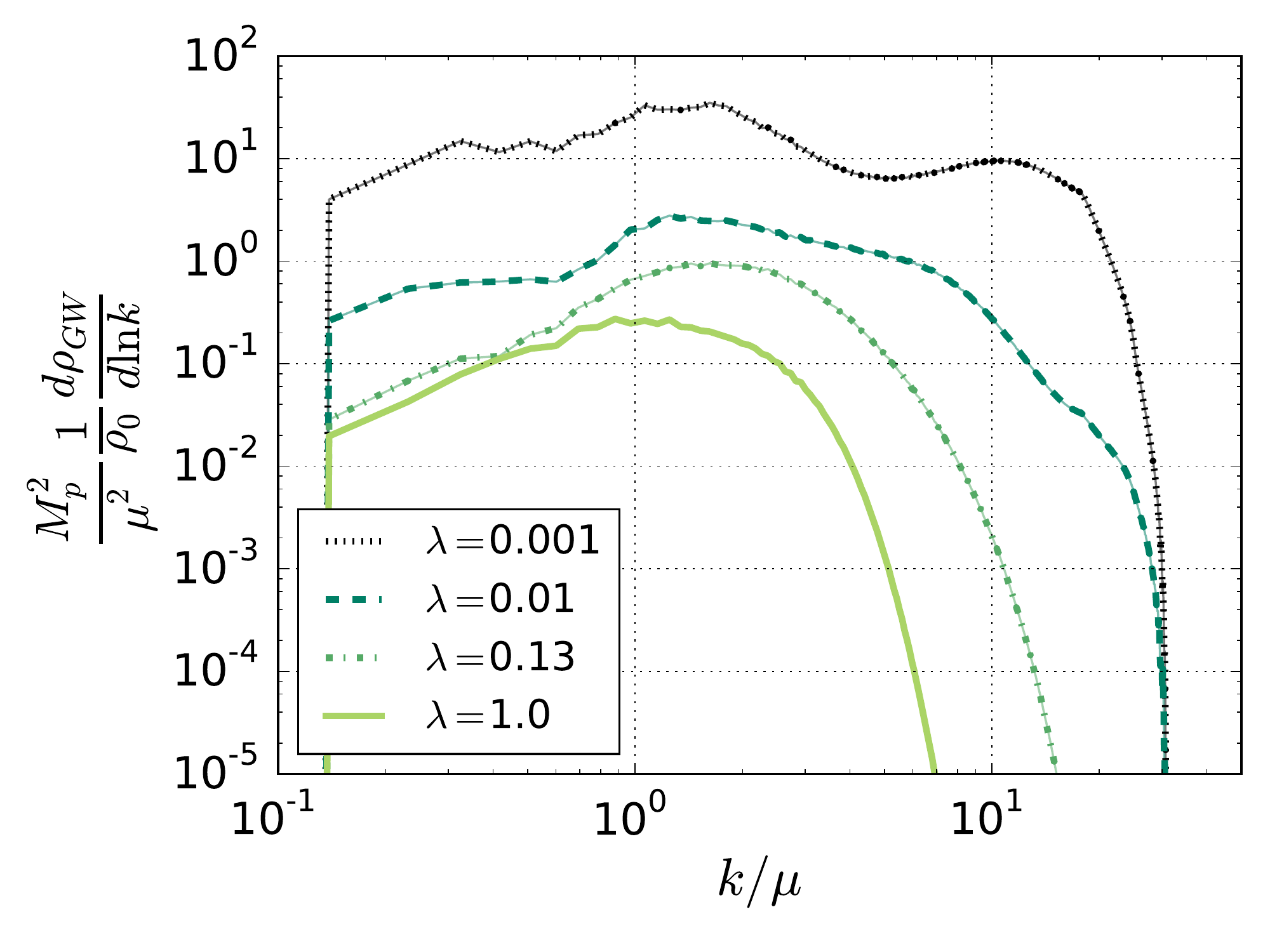}
\includegraphics[width=0.5\textwidth]{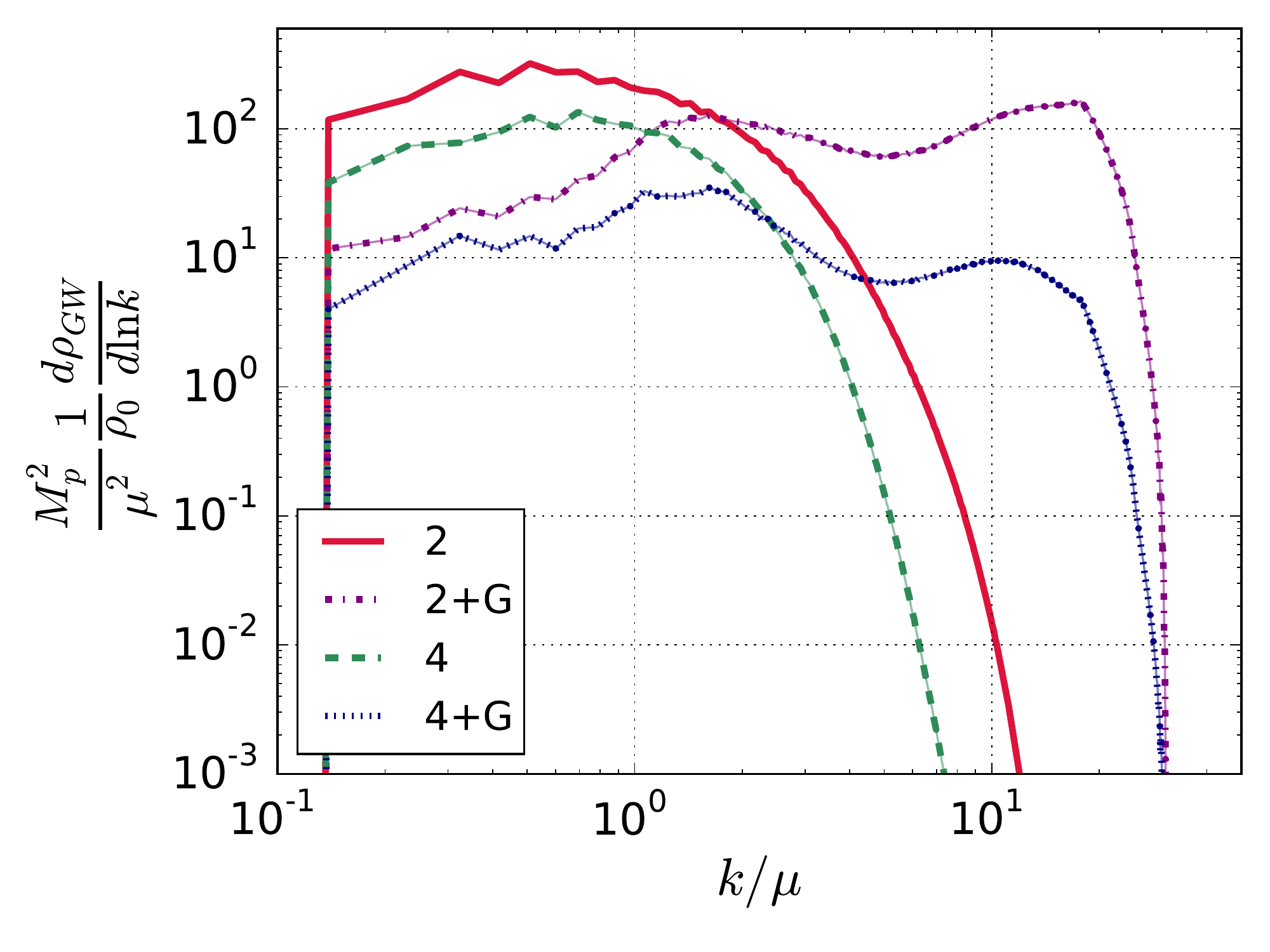}
\caption{The SU(2)-Higgs spectrum for different $\lambda$ (left), and
  for $\lambda=0.001$ for all four models (right). $\mu\tau_q=0$. Note
  the Brillouin zone edge at $k/\mu \approx 18.5$.}
\label{fig:varlam}
\end{figure}

One way to disentangle the two scales is to make $\lambda$ (and hence
$m_H/m_W$) smaller. We will keep $\mu$ fixed, and so what changes is
$m_W$ (in physical and lattice units) and the Higgs vev $v$. In
Fig.~\ref{fig:varlam} (left) we show the spectrum of gravitational
waves for four different values of $\lambda$, for the gauge-scalar
model only. We see that as the gauge boson mass increases ($\lambda$
decreases) the peak resolves into two distinct peaks. The amplitude
also increases significantly, by one or two orders of magnitude.

For $\lambda=0.001$, the W-mass is clearly at the edge of our
dynamical range ($am_W\simeq 7.3 \times am_H$), where lattice artefacts dominate, and we can therefore
not go to even smaller Higgs coupling. In order to confirm the origin
of the second peak, we show in Fig.~\ref{fig:varlam} (right) all four
models with $\lambda=0.001$. There is now no trace of the second peak,
showing that the gauge field is the cause of it (and not, say, the
value of $v$). The increase in magnitude as $\lambda$ is decreased is
common for all the models. For smaller $\lambda$, the exponential
tachyonic instability lasts longer.

Such peak structure and other tell-tale features of multiple
mass-scales would be possible targets for observations. However, the
Higgs and W-mass scales are unfortunately far from the observational
range of LISA. Tuning the parameters to shift the peaks into that range, although perhaps possible in principle, is not our main interest here. 

\section{Discussion and conclusion}
\label{sec:discussion}

As mentioned, we take the energy scale $\mu$ small enough relative to
the Planck mass that we can ignore the expansion of the Universe
during the time-scale of the simulation. Then we have the simple
relation between a given physical scale on the lattice $ak$ and the
frequency $f$ today \cite{Dufaux:2007pt}
\begin{eqnarray}
f= 4\times 10^{10}\textrm{ Hz} \left(
\frac{ak}{a\rho^{1/4}}
\right)
= 4\times 10^{10}\textrm{ Hz} \left(
\frac{k}{\mu}\right)(4\lambda)^{1/4}
=3.4\times 10^{10}\textrm{ Hz} \times \frac{k}{\mu},
\end{eqnarray}
where we have taken the value $\lambda=0.13$ (multiply by $0.3$ for
$\lambda=0.001$). Hence, for fixed $\lambda$, the peak frequency for
any choice of $\mu$ can be read off from the figures.  Similarly, the
amplitude of the spectrum is given by
\begin{eqnarray}
\Omega_{\rm gw}h^2 = \frac{1}{\rho}\frac{d \rho_\text{GW}}{d\ln k}\left(\frac{g_*}{g_0}\right)^{-1/3}\Omega_{\rm rad}h^2=9.3\times 10^{-6}\times \frac{1}{\rho}\frac{d \rho_\text{GW}}{d\ln k},
\end{eqnarray}
using $\Omega_{\rm rad}h^2=4.3\times 10^{-5}$ and $g_*/g_0\simeq
100$. Again, this amplitude may therefore be read off from the
figures, remembering to rescale by $\mu^2/M_{\rm p}$.

The peak sensitivity of the LISA mission is around $0.01$ Hz, whereas
our maximum signal occurs at $k/\mu\simeq 1.5$, corresponding to
$5\times 10^{10}$ Hz. The peak amplitude is
\begin{eqnarray}
  \label{eq:finalamp}
\Omega_{\rm gw}h^2 = 9.3\times 10^{-6}\left(\frac{\mu}{M_{\rm p}}\right)^2,
\end{eqnarray}
which is $10^{-38}$ for the electroweak scale and $10^{-12}$ for a
GUT-scale transition. This applies to $\lambda=0.13$, and we have seen
that a few orders of magnitude can be gained by decreasing
$\lambda$. Increasing the quench time decreases the magnitude of the
gravitational spectrum, once the quench is slower than the finite
time-scale off the Higgs roll-off.

When including gauge fields, we observe a stronger suppression in the
IR (see also Ref.~\cite{Dufaux:2010cf} for the Abelian case) with a
power-law slope of about $1.6$ near the peak. This gives way to a
near-linear dependence further from the peak, although at very long
wavelengths the causal behaviour of the source implies a steeper cubic
power law. Hence, we expect that the signal 13 decades into the IR
will be completely undetectable by LISA.

The most interesting effect of non-Abelian gauge fields is however, that the amplitude of GW decreases relative to a scalar-only theory. The opposite is the case for Abelian gauge fields. Most likely, this is an result of the non-Abelian self-interactions damping out the gauge-field sources of GW.

We saw that varying the coupling $\lambda$, a second peak
corresponding to the gauge field mass emerges. It is possible that
allowing for a very small (or zero) gauge field mass could overcome
the IR suppression, by effectively shifting the gauge field peak far
into the IR. The obvious candidate for this is the Standard Model
photon but fields with very small masses are difficult to contain on a
finite lattice. Also, although the identity of the photon is ambiguous
during the tachyonic transition, once the Higgs mechanism is realised,
the mass really is zero. There is no parameter with which to gradually
``turn the mass off''. We did not implement the photon in our
simulations, and postpone a resolution of this issue to future work.

The peak signal frequency depends only on the scale $\mu$ through the
combination $k/\mu$, because we assume that the Hubble rate is
determined by the Standard Model-inflaton energy component only. If
another energy component than the Higgs potential would dominate the
expansion of the Universe, the peak would redshift
differently. Introducing such a new component, one would have to
account for how it decays into SM degrees of freedom prior to BBN.

The hot plasma present in the early universe after reheating is an
additional source of gravitational waves~\cite{Ghiglieri:2015nfa}. In
fact, it might even prove to be a more significant source than the
signal predicted for a tachyonic transition. The amplitude of
gravitational waves from the plasma would be today (for typical
Standard Model values of the energy density and shear viscosity)
\begin{equation}
\Omega_\text{gw} h^2 \approx 10^{-6}
\left(\frac{T_\text{max}}{M_{\rm p}}\right)
\end{equation}
where $T_\text{max}$ would be the maximum temperature of the plasma,
as it forms. This can be parametrically larger than the contribution
of Eq.~(\ref{eq:finalamp}) by a factor $M_{\rm p} T_\text{max}/\mu^2$,
with a peak at wavenumber $k \sim T_0$, the temperature at which
electroweak symmetry breaking takes place. This is in contrast to a
first-order phase transition, where such plasma dynamics will be an
insignificant source of gravitational waves except at the highest
frequencies.

It is not a priori unreasonable to think that the copious production
of particles in the IR would allow for detection of a tachyonic
transition through gravitational waves. We have however seen that
including non-Abelian gauge fields further suppresses the signal
relative to scalar-only theories. With the possible caveats described
above we must conclude that tachyonic preheating will not be
observable at LISA.

\vspace{0.2cm}

\noindent
{\bf Acknowledgments:} AT is supported by a UiS-ToppForsk grant from
the University of Stavanger. ST is supported by the Magnus Ehrnrooth
foundation and Academy of Finland grant 267842. DJW is supported by
Academy of Finland grant 267286. The authors thank Mark Hindmarsh for
enlightening discussions.  The numerical work was performed on the
Abel Cluster, owned by the University of Oslo and the Norwegian
metacenter for High Performance Computing (NOTUR), and operated by the
Department for Research Computing at USIT, the University of Oslo
IT-department.

\bibliographystyle{jhep.bst}

\end{document}